\documentclass{interact}
\let\cedilla\c

\newcommand{\ii}{\ensuremath{\mathrm{i}}}
\newcommand{\ee}{\ensuremath{\mathrm{e}}}
\newcommand{\dd}{\mathrm{d}}

\newcommand{\bra}[1]{\ensuremath{\left\langle #1 \right\rvert}}
\newcommand{\ket}[1]{\ensuremath{\left\lvert #1 \right\rangle}}

\newcommand{\ketbra}[2]{\ensuremath{\left| #1 \middle\rangle\middle\langle #2\right|} }

\newcommand{\braket}[2]{\ensuremath{\left\langle #1 \vphantom{#2} \middle\vert  #2 \vphantom{#1}\right\rangle}}

\renewcommand{\Im}{\,\mathrm{Im}\,}

\newcommand{\Tr}{\mathrm{Tr}}

\renewcommand{\c}{\ensuremath{\hat{c}}}

\usepackage[T1]{fontenc}
\usepackage{graphicx,epsfig}
\usepackage{epstopdf}
\usepackage{txfonts}
\usepackage{mathrsfs}
\usepackage{amsmath,amssymb,bbm,bm}
\usepackage{float}
\usepackage{import}
\usepackage{booktabs}
\usepackage{exscale}
\usepackage[english]{babel}
\usepackage{color}
\usepackage{appendix}
\usepackage{dsfont}
\usepackage{hyperref}
\hypersetup{
    colorlinks,
    linkcolor={rgb:red,2; blue,1},
    citecolor={blue!60!black},
    urlcolor={blue!60!black}
}
\usepackage{xcolor}

\usepackage[numbers,sort&compress]{natbib}
\bibpunct[, ]{[}{]}{,}{n}{,}{,}
\makeatletter
\def\NAT@def@citea{\def\@citea{\NAT@separator}}
\makeatother

\begin{document}

\title{Quantum thermal absorption machines: refrigerators, engines and clocks}

\author{
\name{Mark T. Mitchison\thanks{markTmitchison@gmail.com}}
\affil{School of Physics, Trinity College Dublin, College Green, Dublin 2, Ireland}
}

\maketitle

\begin{keywords}
Quantum thermodynamics, open quantum systems.
\end{keywords}

\begin{abstract}
The inexorable miniaturisation of technologies, the relentless drive to improve efficiency and the enticing prospect of boosting performance through quantum effects are all compelling reasons to investigate microscopic machines. Thermal absorption machines are a particularly interesting class of device that operate autonomously and use only heat flows to perform a useful task. In the quantum regime, this provides a natural setting in which to quantify the thermodynamic cost of various operations such as cooling, timekeeping or entanglement generation. This article presents a pedagogical introduction to the physics of quantum absorption machines, covering refrigerators, engines and clocks in detail. 
\end{abstract}

\section{The appeal of autonomy in quantum thermodynamics}
\label{sec:intro}

Quantum thermodynamics aims to describe how quantum-mechanical laws affect our ability to manipulate energy at the nanoscale. This knowledge deepens our understanding of the foundations of thermodynamics but it might also be applied, for example, to reduce the power consumption of quantum technologies or to harvest energy from tiny, ubiquitous temperature gradients. There is also the exciting possibility of harnessing quantum resources like entanglement to carry out such tasks in qualitatively different and better ways. Regardless of such possible advantages, the rapid miniaturisation of technologies ranging from consumer electronics to medical sensing demands a profound comprehension of the energetic constraints imposed by quantum mechanics. Together, these motivations are inspiring a new generation of experiments in controlled quantum dynamics and spurring the development of better theoretical methods to tackle quantum physics far from equilibrium.

One of the essential insights of classical thermodynamics is that different forms of energy transfer can be classified according to how useful they are. This is encapsulated by the first law of thermodynamics, which distinguishes work from heat. A principal conceptual challenge in the field of quantum thermodynamics is to extend ideas such as heat and work to the quantum domain. Clearly, the notion of mechanical work is not necessarily applicable to a quantum system like a magnetic spin or an optical field. Instead, one may appeal to the more general definition of work in statistical mechanics, which is a change in internal energy due to the variation of some external control parameters that modify the Hamiltonian of the system. More precisely, if $U = \sum_n p_n E_n$ denotes the internal energy of a system whose energy levels $E_n$ are occupied with probabilities $p_n$, then the change in internal energy is
\begin{align}\label{internal_energy_change}
\dd U = \underbrace{\sum_n  p_n\dd E_n}_{\rm work} + \underbrace{\sum_n  E_n\dd p_n}_{\rm heat}.
\end{align}
For example, applying an external magnetic field performs work on a quantum spin by increasing the difference between its energy levels. On the other hand, a thermal bath induces random spin flips, which change the occupation probabilities of the energy levels and thus constitute heat transfer. 

From this viewpoint, a quantum heat engine performs work by transferring energy to an external, classical system. This framework is mathematically convenient, because the work is defined purely in terms of the variables $p_n$, $E_n$ describing the quantum system. However, it becomes conceptually problematic once we try to precisely quantify the true power and efficiency of energy conversion in the quantum realm. This is important because the tiny energy budget of quantum machines calls for very careful accounting. In particular, the work that can be extracted from a quantum system, such as a single atom, typically pales in comparison to the macroscopic quantities of energy needed to maintain the control fields, such as laser beams, used to extract said work. Indeed, for the external system to be well approximated as classical, it must remain for all intents and purposes unaffected  by the quantum system to which it is coupled. This raises the question of whether it is even meaningful, let alone practically useful, to talk about a quantum system performing work on a classical one. 

The resolution of the paradox is, of course, to acknowledge that Nature is quantum, not classical. Ideally, therefore, a complete description of a quantum thermal machine should explicitly incorporate the quantum dynamics of all sources of work and external control, as well as the resources needed to prepare them. This leads us to consider \textit{autonomous} thermal machines, which operate under a time-independent Hamiltonian and use heat currents to perform a useful task. Autonomous thermal machines provide a framework to consistently quantify the energetic cost of different operations in terms of a universal currency, namely heat exchanged with thermal reservoirs. This represents the ``lowest common denominator'' of energetic resources because thermal states are naturally abundant, easy to prepare and cost little to maintain by virtue of the tendency of macroscopic systems to equilibrate. Quantifying energy expenditure purely in terms of heat helps to prevent hidden costs --- such as those associated with preparing macroscopically coherent driving fields --- from sneaking into the analysis unaccounted for.

The aim of this review is to explore the physics of autonomous thermal machines by way of pedagogical example. Our focus is restricted to continuous (i.e. non-reciprocating) devices where energy is the only relevant conserved quantity. For want of a better term, such devices are referred to here as thermal absorption machines. For example, quantum absorption refrigerators are machines that cool using only a hot bath as an energy source. In the following, we discuss how energy quantisation and quantum coherence affect the performance of absorption refrigerators, and how the finite energy and size of quantum systems limits the efficiency of heat-to-work conversion in absorption engines. We also show that absorption machines provide a natural setting to discuss the thermodynamic limitations of measuring time. Since experimental investigations of truly autonomous absorption machines in the quantum regime are extremely challenging, the majority of work done in this area has focused on theory and therefore so does this review. Nevertheless, we connect the theoretical results to experiments and potential physical implementations, where possible. 

As outlined above, the scope of this article is rather modest compared to the many excellent reviews of quantum thermodynamics that can be found in the literature. Topical surveys of the entire field can be found in Refs.~\cite{Vinjanampathy2016} and, more recently, in Ref.~\cite{Binder2019}. A comprehensive review of thermoelectric machines is provided in Ref.~\cite{Benenti2017}. For reasons of concision and cohesion, we regretfully do not discuss the fascinating topic of autonomous reciprocating engines; an overview of recent progress in this area can be found in Ref.~\cite{Seah2018b}. The connections between (quantum) information theory and thermodynamics are expounded in Refs.~\cite{Parrondo2015,Goold2016}.  Refs.~\cite{Kosloff2013,Kosloff2014} review the dynamics of quantum thermal machines in technical detail. Finally, the seminal book by Gemmer et al.~\cite{Gemmer2004} discusses key concepts in the thermodynamics of composite quantum systems. 

The remainder of the text is structured as follows. Sec.~\ref{sec:open_systems} introduces a few basic elements of the theory of open quantum systems that are necessary to understand this review. Sec.~\ref{sec:QAR} covers quantum absorption refrigerators, which are in a sense the template for all continuous absorption machines. Hence, many of the results here apply to later sections as well. In Sec.~\ref{sec:engines} we turn to absorption engines, focusing on ergotropy change as a measure of work. In Sec.~\ref{sec:clocks}, we discuss thermodynamic constraints on the measurement of time using absorption machines. A brief summary and outlook on future directions is given in Sec.~\ref{sec:summary}. We make use of ``natural quantum-thermodynamic units'' throughout, in which both the Planck and Boltzmann constants are equal to one.

\section{Absorption machines as open quantum systems}
\label{sec:open_systems}

\subsection{The quantum master equation}
\label{sec:master_equation}

A quantum thermal absorption machine generates a useful output by harnessing the flow of heat from (or into) at least two thermal reservoirs. It goes without saying, therefore, that the behaviour of the device is strongly influenced by its environment, i.e. it is an open quantum system. Generally speaking, coupling to a macroscopic environment introduces an element of stochasticity into the evolution of open quantum systems, which must therefore be described by a statistical mixture of pure states. Furthermore, quantum absorption machines necessarily have non-linear internal dynamics and operate under highly non-equilibrium conditions. This makes understanding and predicting their behaviour quite challenging, despite their small size. 

Nevertheless, various theoretical techniques do exist for dealing with such problems. Here, we focus on one of the most popular and versatile tools: the quantum master equation. This is appropriate for dealing with situations in which the open system interacts weakly with a Markovian (i.e. memoryless) environment. An advantage of this approach is that interactions within the machine are captured exactly, albeit at the cost of treating the coupling to the baths perturbatively.

Before delving into a detailed discussion of the master equation, let us first recall some elementary facts about quantum mixed states.  In many situations of interest, we are ignorant of the exact wave function of the system and must instead ascribe probabilities to the different possible states that the system may have. In such cases, the system is described by a density operator $\rho$, which may be diagonalised to give
\begin{equation}\label{mixed_state}
\rho = \sum_n p_n \ketbra{\psi_n}{\psi_n}.
\end{equation}
Each eigenvalue $p_n$ gives the probability to find the system in the corresponding pure state $\ket{\psi_n}$, where the set of eigenvectors $\{\ket{\psi_n}\}$ forms a complete, orthonormal basis. For example, a system in thermal equilibrium at temperature $T =1/\beta$ occupies each energy eigenstate $\ket{E_n}$ with probability given by the Boltzmann distribution $p_n = \ee^{-\beta E_n}/Z$, where $E_n$ is the corresponding energy eigenvalue and $Z = \sum_n \ee^{-\beta E_n}$ is the partition function. The density operator is a complete description of the system, since it allows us to predict the expectation value of any observable $\hat{O}$ via the trace formula $\langle \hat{O}\rangle = \Tr \{\hat{O}\rho\} = \sum_n p_n \langle \psi_n|\hat{O}|\psi_n\rangle$. In a bipartite system described by a state $\rho_{AB}$, all observables of one part, say $A$, can be found from the reduced density matrix $\rho_A = \Tr_B\{\rho_{AB}\}$, where $\Tr_B$ denotes a partial trace over the Hilbert space of $B$.

In a closed quantum system, pure states evolve in time according to the Schr\"odinger equation $\partial\ket{\psi} /\partial t = -\ii \hat{H}\ket{\psi}$, where $\hat{H}$ is the Hamiltonian operator. For mixed states, this generalises to the von Neumann equation
\begin{equation}\label{von_Neumann}
\dot{\rho} = -\ii[\hat{H},\rho],
\end{equation}
where the dot denotes a partial time derivative. The solution of this equation is 
\begin{equation}\label{unitary_channel}
\rho(t) = \mathcal{U}[\rho(t_0)] =  \hat{U} \rho(t_0) \hat{U}^\dagger ,
\end{equation}
where $\hat{U} = \ee^{-\ii \hat{H}(t-t_0)}$ is the unitary time-evolution operator. Here and henceforth we use a calligraphic letter such as $\mathcal{U}$ to denote a ``super-operator'', i.e. a map between density operators. Applying the above transformation to Eq.~\eqref{mixed_state}, we see that unitary evolution rotates each eigenvector of the density matrix to a new configuration $\ket{\psi_n(t)} = \hat{U}\ket{\psi_n}$ but leaves the corresponding probabilities $p_n$ unchanged. 

In an open quantum system, however, the interaction with the environment generally changes the occupation probabilities of the system density matrix. For example, an ideal thermal bath drives the system towards the Boltzmann distribution, irrespective of the initial state.  Indeed,  although the system-environment density matrix $\rho_{ S\!E}$ evolves unitarily, the reduced density matrix $\rho(t) = \Tr_E\{\hat{U}\rho_{ S\!E}(t_0) \hat{U}^\dagger\}$ may not. If the system and bath are initially uncorrelated, so that $\rho_{S\! E}(t_0) = \rho(t_0) \otimes \rho_E(t_0)$, the system density matrix undergoes a more general kind of transformation
\begin{equation}\label{quantum_channel}
\mathcal{E}[\rho] =  \sum_\mu \hat{K}_\mu \rho \hat{K}_\mu^\dagger,
\end{equation}
which is known in the literature as a quantum operation~\cite{Nielsen2009}. The so-called Kraus operators $\hat{K}_\mu$ must obey the constraint $\sum_\mu \hat{K}_\mu^\dagger \hat{K}_\mu = \mathbbm{1}$. This condition ensures that the operation is completely positive and trace-preserving (CPTP), which means that the eigenvalue spectrum of the density operator remains positive and normalised.\footnote{\textit{Complete} positivity means that the operation $\mathcal{E}\otimes \mathcal{I}$ is positive, where $\mathcal{I}$ is the identity superoperator acting on an arbitrary auxiliary system.} Note that unitary evolution is a special case of Eq.~\eqref{quantum_channel} in which there is only a single Kraus operator. 

A differential equation that generates a non-unitary CPTP map is known as a quantum master equation. It is obtained simply by adding an extra term to the right-hand side of the von Neumann equation~\eqref{von_Neumann}, viz.
\begin{equation}\label{master_equation}
\dot{\rho} = -\ii [\hat{H},\rho] + \mathcal{L}[\rho].
\end{equation}
The non-unitary effect of the environment is encapsulated by a super-operator $\mathcal{L}$ known as the dissipator. In general, the environment may also modify the Hamiltonian, but at weak system-bath coupling such effects are often negligible. 

Finding the exact dissipator is generally an intractably hard task, which would require a complete solution to the many-body problem of the environment interacting with the system. It is thus necessary to resort to simplifying approximations that capture the relevant physics in an effective way. In particular, the Born-Markov approximation is based on the assumption that the system and environment remain weakly correlated and thus perturbations made by the system on the environment do not significantly affect the system's future evolution. This is genereally a good assumption for weak system-bath coupling. The Born-Markov approximation leads to a dissipator that is independent of time, so that the solutions of Eq.~\eqref{master_equation} are related as $\rho(t+\Delta t) = \mathcal{E}_{\Delta t}[\rho(t)]$, where $\mathcal{E}_{\Delta t}$ is a CPTP map that depends only on the time interval $\Delta t>0$.

The most general time-independent dissipator generating a CPTP evolution was found by Gorini, Kossakowski, Sudarshan and Lindblad~\cite{Gorini1976,Lindblad1976}. This is the so-called Lindblad form
\begin{align}\label{Lindblad_form}
\mathcal{L}[\rho] & = \sum_\mu \left (\hat{L}_\mu \rho \hat{L}_\mu^\dagger - \tfrac{1}{2}\{ \hat{L}^\dagger_\mu \hat{L}_\mu,\rho \} \right ),
\end{align}
where each jump operator $\hat{L}_\mu$ describes an environment-induced transition (i.e. a transfer of probability from one system state to another). In general, Lindblad dissipators may either be derived from a microscopic model under certain approximations, or simply postulated on phenomenological grounds. Examples are discussed in the following section and further details may be found in dedicated textbooks such as Refs.~\cite{BreuerPetruccione,Rivas2012}.

\subsection{Thermodynamics from the master equation}
\label{sec:thermal_ME}

Consider the dynamics of an open quantum system immersed within a thermal bath. Assuming weak system-bath coupling, the asymptotic state at long times is a canonical equilibrium state
\begin{equation}\label{equilibrium_state}
\rho = \frac{\ee^{-\beta \hat{H}}}{Z}.
\end{equation}
One easily checks that this is a stationary state of Eq.~\eqref{master_equation}, i.e. $\dd \rho /\dd t = 0$, so long as $\mathcal{L}[\ee^{-\beta \hat{H}}] = 0.$ Such a dissipator can be derived from a microscopic system-bath Hamiltonian under the Born-Markov and secular approximations~\cite{BreuerPetruccione,Rivas2012}. The resulting physical picture is that transitions between pairs of energy eigenstates, say $\ket{E_{k}}\to \ket{E_{l}}$, are driven by exchanging quanta of frequency $\omega = E_l -E_k$ with the bath. More precisely, the dissipator takes the form\footnote{Strictly speaking, there could be multiple jump operators $\hat{A}_j(\omega)$ and rates $\gamma_j(\omega)$ for each transition frequency $\omega$, but we do not need to consider such complications here.}
\begin{equation}\label{detailed_balance_dissipator}
\mathcal{L}[\rho] = \sum_{\omega} \gamma(\omega)\left (\hat{A}(\omega) \rho \hat{A}^\dagger(\omega) - \tfrac{1}{2}\{ \hat{A}^\dagger(\omega) \hat{A}(\omega),\rho \} \right ) ,
\end{equation}
where the sum runs over all differences $\omega$ between the eigenvalues of $\hat{H}$. The operators $\hat{A}(\omega)$ are lowering operators, i.e.\ they satisfy $[\hat{H},\hat{A}(\omega)] = -\omega \hat{A}(\omega).$ Therefore, if $\omega$ is positive (negative) then $\hat{A}(\omega)$ describes a transition that decreases (increases) the open system's energy due to the emission (absorption) of a quantum of frequency $\omega$. The emission and absorption rates are related by the detailed balance condition
\begin{equation}
\label{detailed_balance_rates}
\frac{\gamma(-\omega)}{\gamma(\omega)} = \ee^{-\beta \omega}.
\end{equation}
 The precise functional form of $\gamma(\omega)$ is not overly important for much of the following discussion. However, it is useful to note that a multi-mode bosonic environment  in $D$ spatial dimensions, such as the electromagnetic field, leads to a power-law frequency dependence $\gamma(\omega)\sim \omega^D$.
 
In cases where multiple, independent thermal environments couple to the open system, we may simply add their contributions together as
\begin{equation}\label{additive}
\mathcal{L} = \sum_\alpha \mathcal{L}_\alpha,
\end{equation}
where $\mathcal{L}_\alpha$ represents the individual effect of environment $\alpha$. Combining this with the master equation~\eqref{master_equation}, we obtain the continuity equation for energy 
\begin{equation}\label{energy_balance}
\frac{\dd }{\dd t}\langle \hat{H}\rangle  = \sum_\alpha J_\alpha, 
\end{equation}
where $J_\alpha = \Tr\{\hat{H}\mathcal{L}_\alpha [\rho]\}$ is the heat current flowing into the system from each bath. Eq.~\eqref{energy_balance} is the first law of thermodynamics in differential form for a thermal absorption machine, whose internal energy changes exclusively due to heat, i.e. no work is done on the system as a whole. The second law of thermodynamics in this context reads as
\begin{equation}\label{second_law_full}
\frac{\dd }{\dd t}S(\rho) - \sum_\alpha \beta_\alpha J_\alpha \geq 0,
\end{equation}
where $S(\rho) = -{\rm Tr} \{\rho \ln \rho\}$ is the von Neumann entropy of the open system and $-\beta_\alpha J_\alpha$ is the rate of entropy change in the macroscopic bath, assuming that its temperature $T_\alpha = 1/\beta_\alpha$ remains constant (because its heat capacity is very large). While the first law~\eqref{energy_balance} is automatically implied by Eqs.~\eqref{master_equation} and \eqref{additive}, the second law~\eqref{second_law_full} represents a non-trivial constraint that a physical evolution should satisfy, which dictates the fundamental performance limits on quantum thermal machines.  

However, in the multi-bath case, thermodynamical considerations alone are not sufficient to fix the form of the dissipator, which also depends on the structure of the open system and the energy and temperature regime of interest. In general, thermal absorption machines are multipartite systems comprising several interacting units that are spatially separated or in some sense distinguishable. This allows the baths to interact with different parts of the machine and thus avoid coupling directly to each other. We may write the Hamiltonian of a multi-partite system generically as
\begin{equation}\label{multipartite_interacting}
\hat{H} = \hat{H}_0 + g \hat{V},
\end{equation}
where $\hat{H}_0 = \sum_j \hat{H}_j$ is the free Hamiltonian, with $\hat{H}_j$ the local energy of subsystem $j$, while $\hat{V}$ describes interactions between the subsystems with an overall energy scale $g$. If $g$ is comparable to or larger than the energy gaps of $\hat{H}_0$, one may use a ``global'' dissipation model where the dissipators take the form~\eqref{detailed_balance_dissipator} and thus obey $\mathcal{L}_\alpha [\ee^{-\beta_\alpha \hat{H}}] = 0,$ i.e.\ each bath tries to bring the \textit{entire} machine to the corresponding temperature. Alternatively, if $g$ is small compared to the energy gaps of $\hat{H}_0$ and the reservoir temperatures, one may use instead a ``local'' dissipation model in which $\mathcal{L}_\alpha$ acts to thermalise only the subsystem to which bath $\alpha$ is connected. It is also possible to use more general Lindblad equations that interpolate between the two approaches, although these usually require additional phenomenological assumptions~\cite{Schaller2008,Kirsanskas2018}.

Recent years have seen a somewhat controversial discussion over the validity of local and global master equations, stimulated by various examples of apparently nonsensical predictions derived from both approaches~\cite{Wichterich2007,Scala2007,Rivas2010,Levy2014,Trushechkin2016,Stockburger2016,Eastham2016,Purkayastha2016,Decordi2017,Hofer2017,Gonzalez2017a,Naseem2018}. In particular, the global model leads to vanishing coherences between energy eigenstates in the steady state, in contradiction with fundamental conservation laws~\cite{Wichterich2007,Salmilehto2012,Mitchison2018}. On the other hand, the local approach has been shown to violate the second law of thermodynamics in certain cases where $[\hat{H}_0,\hat{V}] \neq 0$ even if $g$ is vanishingly small~\cite{Levy2014}, unless a portion of the energy current from each bath is interpreted as work rather than heat~\cite{Barra2015,Strasberg2017,Chiara2018}. In fact, even the assumption of additivity~\eqref{additive} does not necessarily hold in general due to the indirect interaction between the baths via the open system~\cite{Chan2014,Giusteri2017,Mitchison2018,Kolodyinski2018,Maguire2018}. 

Given these inherent limitations, here we avoid promoting one approach in particular. Rather, we take a pragmatic stance and discuss results obtained from different theoretical frameworks within their respective domains of validity, since each one illuminates a different aspect of the physics. The development of more powerful and general methods to describe multipartite, non-linear open quantum systems driven far from equilibrium remains an important challenge for theorists in this field. 

\section{Quantum absorption refrigerators}
\label{sec:QAR}

\subsection{Three-level refrigerator}
\label{sec:three_level_abs}

Cooling is a necessary precursor to many quantum experiments and technologies. Refrigerators are thus a particularly interesting class of quantum absorption machine because they generate an unambiguously useful output using only heat as an energy source. 
The basic aim of refrigeration is to transfer heat from a cold body at temperature $T_c$ to a hot one at temperature $T_h>T_c$. This requires a source of free energy, since the second law of thermodynamics forbids such a process occurring spontaneously. In a power refrigerator, the free energy is in the form of work, e.g.\ mechanical compression of a refrigerant. In an absorption refrigerator, the power source is replaced by a heat bath called the work reservoir at an even hotter temperature $T_w > T_h$. \footnote{Some treatments use the alternative designation $T_h$ for the hottest reservoir and $T_r$ for the intermediate (``room-temperature'') reservoir.}

\begin{figure}
	\centering
	\includegraphics[trim = 50mm 0mm 50mm 0mm, clip, width=0.6\linewidth]{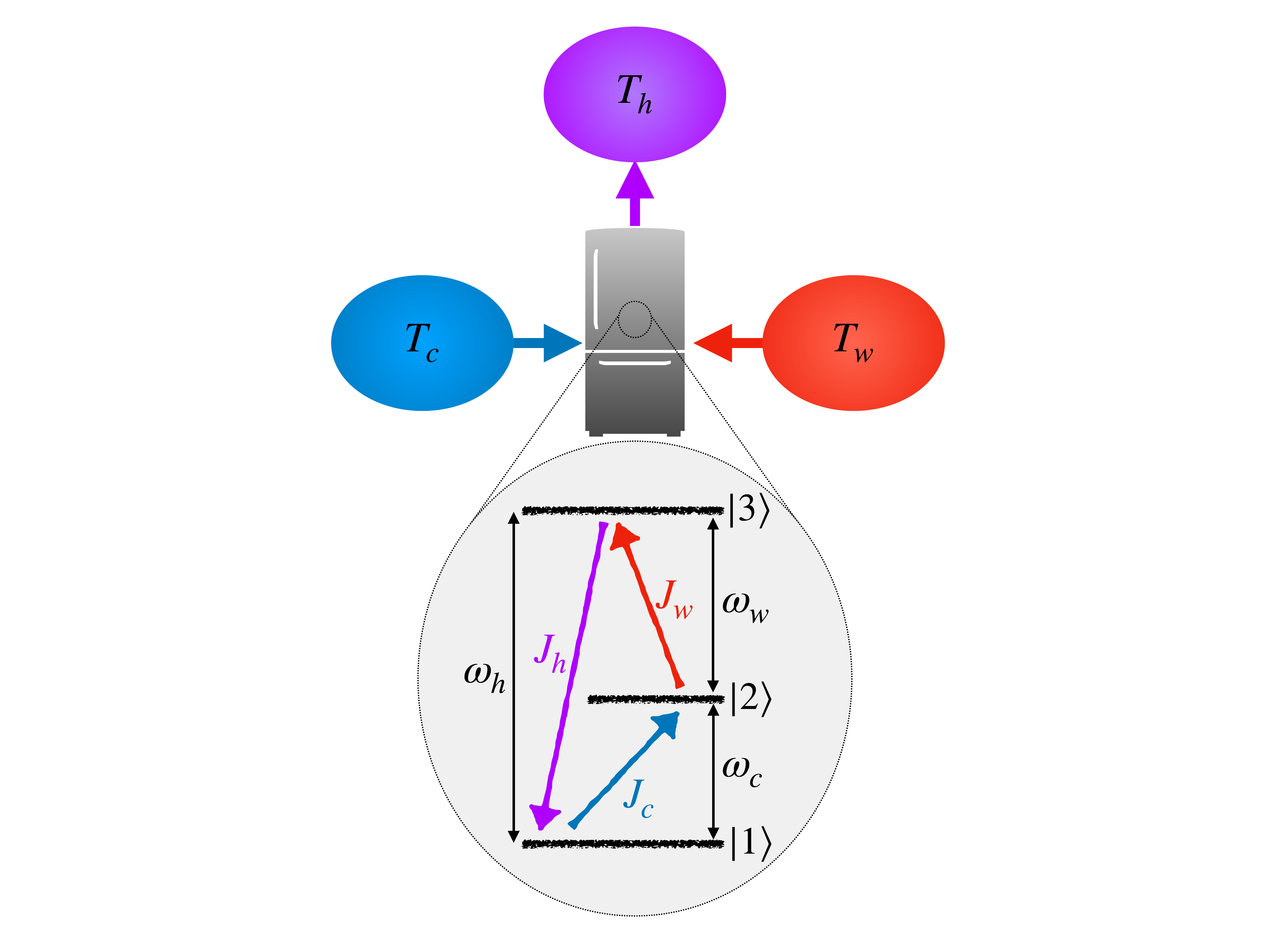}
	\caption{An absorption refrigerator moves heat from a cold to a hot reservoir, powered by a work reservoir at an even higher temperature. The simplest quantum absorption refrigerator comprises three levels with three possible transitions, each coupled to one of the baths.\label{fig:three_level_fridge}}	
\end{figure}

The earliest \textit{quantum} absorption refrigerator to be studied --- and also probably the simplest to understand --- is the three-level model depicted in Fig.~\ref{fig:three_level_fridge}. This set-up dates back to the pioneering work of Scovil and Schulz-DuBois~\cite{Scovil1959}, who modelled the amplification of radiation inside the gain medium of a maser (or laser) as a thermal machine. By running the amplifier in reverse one obtains an absorption refrigerator~\cite{Geusic1967, Palao2001}.

There are three possible transitions between the three states of the system, with energy differences denoted by $\omega_c$, $\omega_h$ and $\omega_w = \omega_h- \omega_c$. Each of these transitions (i.e. each pair of states) is coupled to a different heat reservoir at the corresponding temperature $T_c$, $T_h$ or $T_w$. In order to ensure that each transition couples only to one bath, filters must be used to select only those frequencies in bath $\alpha$ close to resonance with $\omega_\alpha$. 

Let us first assume that the coupling to the cold bath is negligible. The work and hot transitions equilibrate with their respective baths, which implies that $p_3/p_1 =\ee^{-\beta_h \omega_h}$ and $p_3/p_2  = \ee^{-\beta_w \omega_w}$. The population ratio of the cold transition is therefore given by 
\begin{equation}\label{cold_transition_pop}
\frac{p_2}{p_1} = \ee^{\beta_w \omega_w-\beta_h \omega_h } \equiv \ee^{-\beta_v \omega_c},
\end{equation}
where the second equality defines the ``virtual temperature'' $T_v = 1/ \beta_v$, i.e.
\begin{equation}\label{virtual_temp}
 T_v = \frac{\omega_h - \omega_w}{\beta_h \omega_h - \beta_w \omega_w}.
\end{equation}
By a judicious choice of energies and temperatures, we arrange for $T_v$ to be smaller than $T_c$. Now, switching on a finite but weak coupling to the cold reservoir allows excitations to be emitted and absorbed by the transition $\ket{1}\leftrightarrow\ket{2}$. So long as $T_c>T_v$, the cold bath tries to increase the effective temperature of this transition. This leads to heat flow from the cold reservoir into the refrigerator and subsequently out to the hot bath.

The cooling power is $J_c$, the heat current entering from the cold bath. The input power driving the refrigerator is $J_w$, the heat current from the work bath. Hence, the efficiency of cooling is characterised by the coefficient of performance (COP)
\begin{equation}\label{coeff_performance}
\varepsilon = \frac{J_c}{J_w}.
\end{equation}
The COP may be colloquially referred to as an ``efficiency'' even though, unlike an engine's efficiency, a refrigerator's COP may be greater than unity. 

Each quantum of energy $\omega_c$ absorbed from the cold body is accompanied by precisely $\omega_w$ absorbed from the work reservoir and $\omega_h$ dumped in the hot reservoir. It follows that the energy currents must be proportional, i.e. $J_{c,w} = k \omega_{c,w}$ and $J_h = - k \omega_h$ for some constant $k$. Therefore, the COP is given by
\begin{equation}\label{ideal_COP}
\varepsilon = \frac{\omega_c}{\omega_w}.
\end{equation}
This ideal, temperature-independent result expresses the perfect cooperative transfer of precisely one energy quantum each from the cold and work reservoirs into the hot one. The condition for cooling is $T_v < T_c$, which can be written using Eq.~\eqref{virtual_temp} as a ``cooling window'' for the qubit energies
\begin{equation}\label{cooling_window}
0 < \frac{\omega_c}{\omega_w} < \frac{T_c(T_w-T_h)}{T_w(T_h-T_c)}.
\end{equation}
Combining this with Eq.~\eqref{ideal_COP} yields the efficiency bound
\begin{equation}\label{Carnot_COP}
\varepsilon < \varepsilon_{\rm Carnot} = \frac{T_c(T_w-T_h)}{T_w(T_h-T_c)},
\end{equation}
where $\varepsilon_{\rm Carnot}$ is the maximum COP allowed by thermodynamics.  Indeed, the Carnot bound~\eqref{Carnot_COP} is merely an expression of the first and second laws, which, for an absorption refrigerator under steady-state conditions, read as
\begin{align}
\label{first_law_abs}
J_c + J_h + J_w &= 0,\\
\label{second_law_abs}
-\beta_c J_c - \beta_h J_h - \beta_w J_w& \geq 0.
\end{align}
Combining these with the definition~\eqref{coeff_performance} and the assumption that $J_c>0$ leads directly to the bound~\eqref{Carnot_COP}. 

The Carnot limit $\varepsilon \to \varepsilon_{\rm Carnot}$ describes reversible operation, where the entropy production rate vanishes and thus inequality~\eqref{second_law_abs} is saturated. In the language of virtual temperature, this point corresponds to $T_v = T_c$ so that the cooling power is zero. The fact that there is no entropy production at $T_v=T_c$ is rather intuitive, since moving heat between two bodies at the same temperature is thermodynamically reversible. Moreover, the cooling window~\eqref{cooling_window} disappears as $T_c\to 0$, which expresses the impossibility of cooling to absolute zero, i.e. the third law of thermodynamics. Hence, we see the thermodynamic constraints on refrigeration emerging naturally from the dynamics of this simple open quantum system.

\subsection{Three-body refrigerator}
\label{sec:three_body}

The three-level model is important for developing intuition but it lacks several important features of realistic quantum absorption machines. We now discuss a slightly more complicated set-up where each bath couples locally to a separate subsystem. This provides an explicit physical mechanism to filter out all but a certain range frequencies from each bath, and is a natural setting in which to better understand internal sources of irreversibility as well as the role of quantum coherence and entanglement. 

\begin{figure}
	\centering
	\includegraphics[trim =40mm 90mm 40mm 40mm,clip,width=0.6\linewidth]{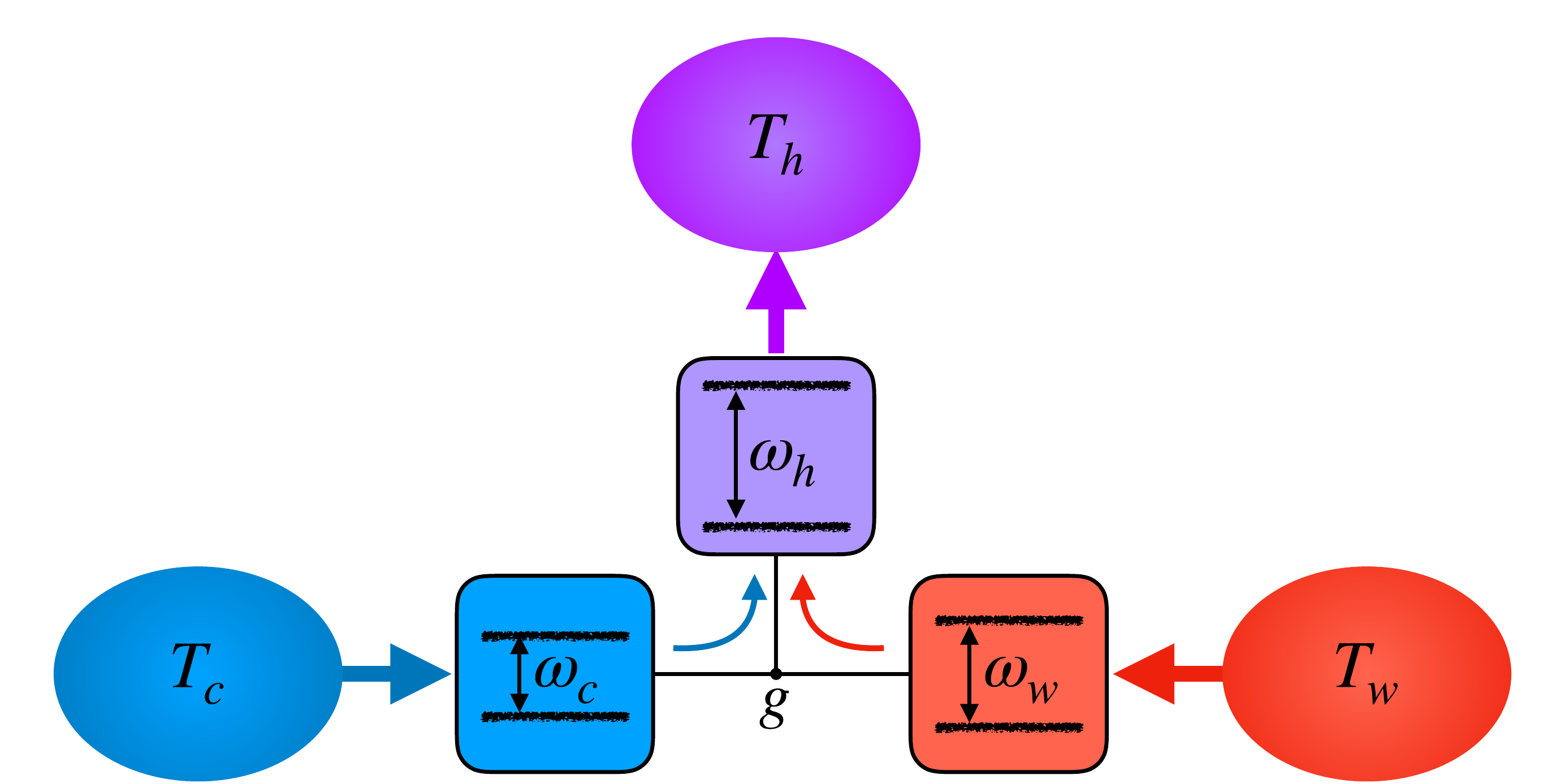}
	\caption{A three-qubit absorption refrigerator cooperatively transfers heat from the cold and work reservoirs into the hot one via a three-body interaction. \label{fig:three_body_fridge}}
\end{figure}

The minimal model of a quantum absorption refrigerator with local system-bath couplings comprises three quantum systems labelled by $\alpha = c,h,w$ coupled to the cold, hot and work reservoirs, respectively~\cite{Linden2010,Levy2012}, as depicted in Fig.~\ref{fig:three_body_fridge}. Each subsystem is required to possess only a single relevant transition frequency, $\omega_c$, $\omega_h$ or $\omega_w$, giving the spacing between adjacent energy levels. This is the case if, for example, these subsystems are two-level systems (qubits)~\cite{Linden2010} or harmonic oscillators~\cite{Levy2012}. The transition frequencies are constrained to obey the condition
\begin{equation}\label{resonanceCondition}
\omega_w  = \omega_h - \omega_c.
\end{equation}
Therefore, an appropriate weak coupling can resonantly transfer energy between the three subsystems.

Such an energy transfer is described by the Hamiltonian
\begin{align}
\label{Htot}
\hat{H} & = \sum_{\alpha=c,h,w} \hat{H}_\alpha + \hat{H}_{\rm int},\\
\label{Halpha}
\hat{H}_\alpha & =  \sum_{n}n\omega_\alpha \ket{n}_\alpha\!\bra{n},\\
\label{three_body}
\hat{H}_{\rm int} & = g \left (\hat{A}_c \hat{A}_h^\dagger  \hat{A}_w  + \hat{A}^\dagger_c \hat{A}_h \hat{A}^\dagger_w \right ).
\end{align}
Here, $\hat{H}_\alpha$ is the Hamiltonian of subsystem $\alpha$ with local eigenstates $\ket{n}_\alpha$, $\hat{H}_{\rm int}$ is the interaction of strength $g$  and $\hat{A}_{\alpha}$ is a lowering operator that obeys the relation $[\hat{H}_\alpha, \hat{A}_\alpha] = -\omega_\alpha\hat{A}_\alpha,$ i.e. it reduces the subsystem's energy by one unit, $\omega_\alpha$. Taking harmonic oscillators for example, $\hat{A}_\alpha = \hat{a}_\alpha$ is the familiar annihilation operator satisfying $[\hat{a}_\alpha,\hat{a}^\dagger_\alpha] = 1$ and $\hat{H}_\alpha = \omega_\alpha \hat{a}^\dagger_\alpha\hat{a}_\alpha$. On the other hand, for qubits described by Pauli operators $\hat{\sigma}^{x,y,z}_\alpha$ we have $\hat{A}_\alpha = \hat{\sigma}^-_\alpha$ and $\hat{H}_\alpha = \omega_\alpha \hat{\sigma}^+_\alpha\hat{\sigma}^-_\alpha$, where $\hat{\sigma}_\alpha^\pm= \tfrac{1}{2}(\hat{\sigma}_\alpha^x \pm \ii \hat{\sigma}_\alpha^y)$. Therefore, the interaction~\eqref{three_body} is such that heat may only flow from $w$ to $h$ by simultaneously removing energy from $c$. Note that this interaction Hamiltonian is trilinear, i.e. it simultaneously couples all three subsystems. For harmonic oscillators, such non-quadratic interactions are necessary to construct an absorption refrigerator~\cite{Martinez2013}. 

\begin{figure}
	\centering
	\includegraphics[trim = 50mm 50mm 50mm 50mm, clip , width=0.5\linewidth]{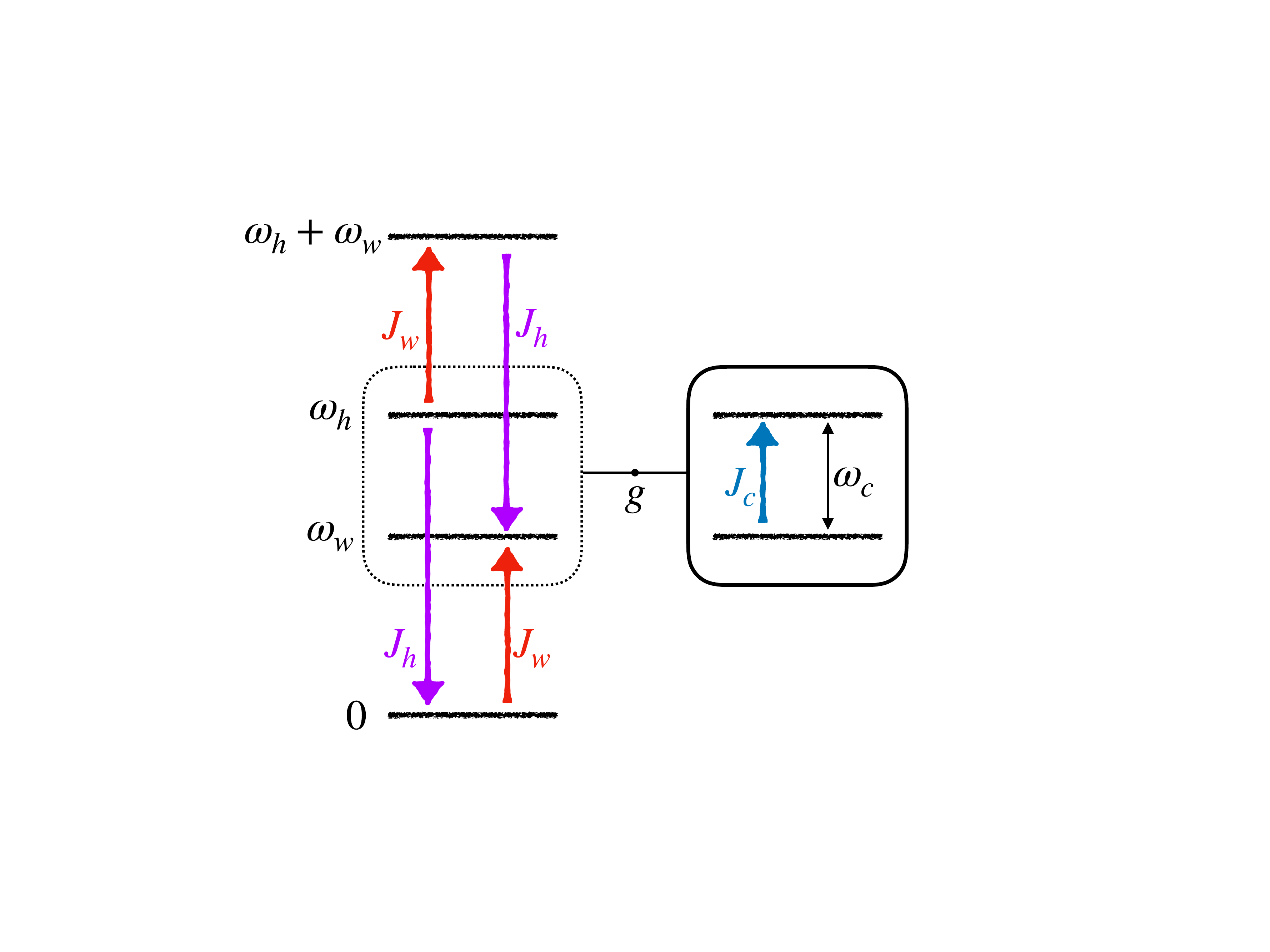}
	\caption{The virtual qubit is a pair of states in the composite Hilbert space of two refrigerator qubits, which couples to the target qubit. Arrows show the direction of transitions driven by the heat currents from the baths. The flow of heat from the work reservoir to the hot one drives the virtual qubit to a very low effective temperature, thus cooling the target. \label{fig:virtual_qubit_fridge}}
\end{figure}

Let us focus now on the three-qubit model introduced by Linden and colleagues~\cite{Linden2010}. In the basis of eigenstates $\{\ket{0}_\alpha$, $\ket{1}_\alpha\}$ of the local Hamiltonians $\hat{H}_\alpha$, with energy eigenvalues $\{0,\omega_\alpha\}$, the interaction Hamiltonian induces transitions between the states
\begin{equation}\label{qubit_fridge_transitions}
 \ket{1}_c \ket{0}_h \ket{1}_w\longleftrightarrow\ket{0}_c\ket{1}_h \ket{0}_w.
\end{equation}
We see that there are just two states of qubits $h$ and $w$ that are relevant for the dynamics of $c$, namely $\ket{0}_v = \ket{0}_h \ket{1}_w$ and $\ket{1}_v= \ket{1}_h \ket{0}_w$. Together these states form a ``virtual qubit''~\cite{Brunner2012} with an energy difference $\omega_h - \omega_w = \omega_c$. Cooling of $c$ is induced by placing the virtual qubit at a colder temperature than $T_c$, as shown in Fig.~\ref{fig:virtual_qubit_fridge}.

To see how this occurs, suppose first that the interaction strength $g$ is negligible, so that qubits $w$ and $h$ equilibrate with their respective baths. The probability $p^{(j)}_\alpha$ that the state $\ket{j}_\alpha$ is occupied is given by $p_\alpha^{(1)}/p_{\alpha}^{(0)} = \ee^{-\beta_\alpha \omega_\alpha}$. Therefore, the virtual qubit states are populated in the ratio
\begin{equation}\label{virtual_qubit_population_ratio}
\frac{p^{(1)}_h p_w^{(0)}}{p^{(0)}_h p^{(1)}_w} = \ee^{\beta_w \omega_w - \beta_h \omega_h}= \ee^{-\beta_v \omega_c},
\end{equation}
where $\beta_v$ is the inverse virtual temperature defined by Eq.~\eqref{virtual_temp}. Switching on the interaction leads to equilibration between the virtual qubit and $c$ via the transitions~\eqref{qubit_fridge_transitions}, or equivalently $\ket{1}_c \ket{0}_v \longleftrightarrow\ket{0}_c\ket{1}_v.$ This swap-like interaction exchanges energy between the cold and virtual qubits. So long as $T_v < T_c$, the predominant direction of energy flow is from $c$ to $v$. The result is cooling of the cold qubit and a net flow of heat from the cold bath into the hot one. 

The notion of virtual qubits and temperatures is a rather convenient and general tool for understanding quantum absorption machines. Refrigerators composed of higher-dimensional subsystems contain many virtual qubits. If these subsystems have more than one transition frequency, then the virtual qubits can also have a distribution of different frequencies (and possibly different virtual temperatures). Thus, in order to cool a complicated system having many different transition frequencies, the machine needs to couple at least one virtual qubit to each transition of the target system. Any mismatch between the virtual temperatures typically leads to a reduction in efficiency, however. 

Other quantum absorption machines can also be analysed using virtual temperatures and qubits. For example, an absorption heat pump features a virtual qubit at a temperature higher than the hottest bath. On the other hand, a heat engine's virtual qubit is at \textit{negative} temperature, leading to population inversion, as discussed in more detail in Sec.~\ref{sec:engines}.

\subsection{Steady-state performance characteristics}
\label{sec:fridge_performance}

The performance of the three-body refrigerator varies significantly depending on the strength of the coupling $g$. In the weak-coupling regime, we use a local thermalisation model to analyse the system. With qubits, for example, the local dissipator $\mathcal{L}_\alpha$ features jump operators $\hat{\sigma}_\alpha^\pm$ and associated gain and decay rates obeying $\gamma^+_\alpha/\gamma^-_\alpha = \ee^{-\beta_\alpha\omega_\alpha}$, so that each qubit is driven separately to equilibrium at the temperature of the corresponding bath. Note that this model is fully consistent with the second law~\cite{Seah2018,Barra2018}, so long as we consider steady-state operation within the cooling window defined by Eq.~\eqref{cooling_window}. 

The steady-state energy currents are proportional to each other at weak coupling~\cite{Skrzypczyk2011,Correa2013, Hofer2016}, i.e.\ $J_{c,w} = k \omega_{c,w}$, $J_h = - k \omega_h$, where the constant of proportionality is 
\begin{align}\label{coherence}
k = 2g \Im \left \langle  \hat{A}_c \hat{A}^\dagger_h \hat{A}_w\right \rangle.
\end{align}
This quantity may be interpreted as the internal current of excitations flowing from $c$ and $w$ into $h$. As a result, the coefficient of performance is given by 
\begin{equation}\label{ideal_COP_3body}
\varepsilon = \frac{\omega_c}{\omega_w},
\end{equation}
indicating that the weak-coupling performance characteristics are identical to those of the three-level refrigerator (c.f.~Eq.~\eqref{ideal_COP}). Therefore, the COP may be adjusted to any value up to the Carnot bound simply by varying the qubit frequencies within the cooling window. The Carnot point corresponding to zero entropy production is obtained at the upper limit of the cooling window, where $T_v = T_c$ so that the cooling power vanishes~\cite{Skrzypczyk2011}. 

The situation is quite different at strong coupling, where the global dissipation model is valid (see Ref.~\cite{Correa2013} for the derivation). As a result, the ideal expression~\eqref{ideal_COP_3body} is no longer correct and one finds that the Carnot COP cannot be achieved even asymptotically. As shown in Ref.~\cite{Correa2015}, the underlying reason for this behaviour is that the interaction Hamiltonian significantly alters the frequency spectrum of quanta emitted and absorbed by the baths. Each of these transition frequencies effectively couples to a different virtual temperature in the machine. Since it is not possible to make all of the virtual temperatures equal to $T_c$ simultaneously, reversible operation cannot be achieved even in principle. 

\begin{figure} 
	\centering
	\includegraphics[trim=80mm 80mm 80mm 80mm, clip, width=0.5\linewidth]{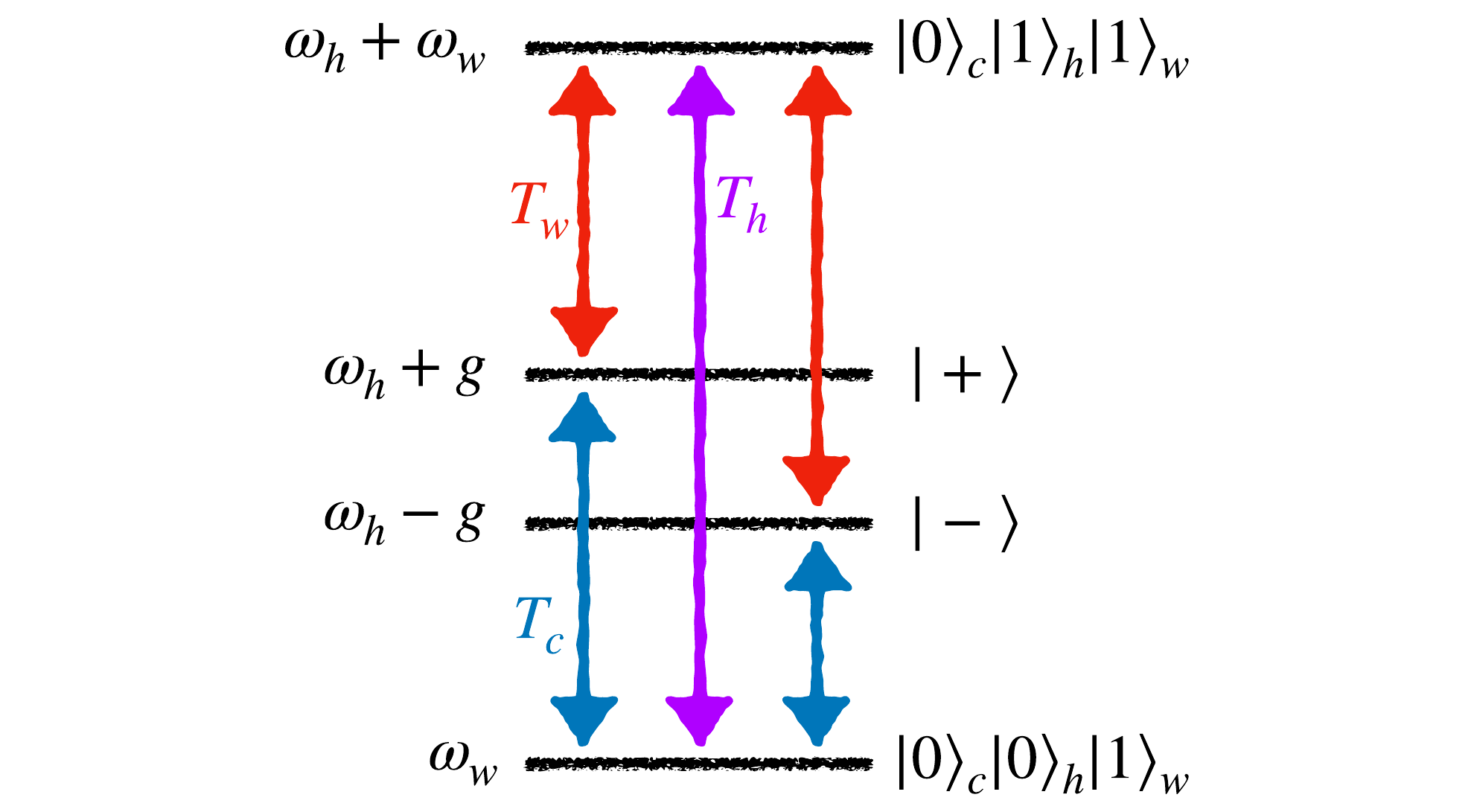}
	\caption{Four eigenstates of the full, interacting Hamiltonian of the three-qubit refrigerator, indicating all possible bath-induced transitions in the global Lindblad approach. The cooling process can be interpreted as a pair of three-level refrigerators running in parallel~\cite{Correa2015}. \label{fig:internal_dissipation}}
\end{figure}

To understand this in more detail, consider the three-qubit case, where the states $\ket{0}_c\ket{1}_h\ket{0}_w$ and $\ket{1}_c\ket{0}_h\ket{1}_w$ are degenerate eigenstates of $\hat{H}_0$ with eigenvalue $\omega_h$. The interaction mixes these states into the symmetric and anti-symmetric combinations
\begin{equation}\label{qubit_fridge_plusminus}
\ket{\pm} = \frac{1}{\sqrt{2}} \left( \ket{0}_c\ket{1}_h\ket{0}_w \pm \ket{1}_c\ket{0}_h\ket{1}_w  \right),
\end{equation}
which are eigenstates of $\hat{H}_0 + \hat{H}_{\rm int}$ with eigenvalues $\omega_h \pm g$. The splitting between these states becomes significant once their energy difference $2g$ is larger than the bath-induced level broadening and comparable to the temperature. In this case, a large number of possible transitions with different frequencies come into play. For example, absorbing a quantum of energy $\omega_c\pm g$ from the cold bath can induce the transitions $\ket{0}_c\ket{0}_h\ket{1}_w \to \ket{\pm}$ or $\ket{\mp}\to \ket{1}_c\ket{1}_h\ket{0}_w$. In Fig.~\ref{fig:internal_dissipation}, we depict just four out of the eight eigenstates showing all the possible bath-induced transitions between them under global Lindblad dissipation. The level scheme resembles a pair of three-level refrigerators running in parallel, with two transitions of frequency $\omega_c\pm g$ coupled to the cold bath. The corresponding virtual temperatures are given by 
\begin{equation}\label{virtual_temps_int_diss}
T_v^\pm = \frac{\omega_c\pm g}{\beta_h\omega_h - \beta_w(\omega_w\mp g)},
\end{equation}
which clearly cannot both be equal to $T_c$. Note that Fig.~\ref{fig:internal_dissipation} also shows the possibility of direct heat exchange between the cold and work baths, which reduces the COP further. However, such heat leaks have a negligible effect on efficiency in comparison to the virtual temperature mismatch. 

\begin{figure}\centering
	\begin{minipage}{0.45\linewidth}
			\flushleft(a)\\
			\centering
		\includegraphics[width=\linewidth]{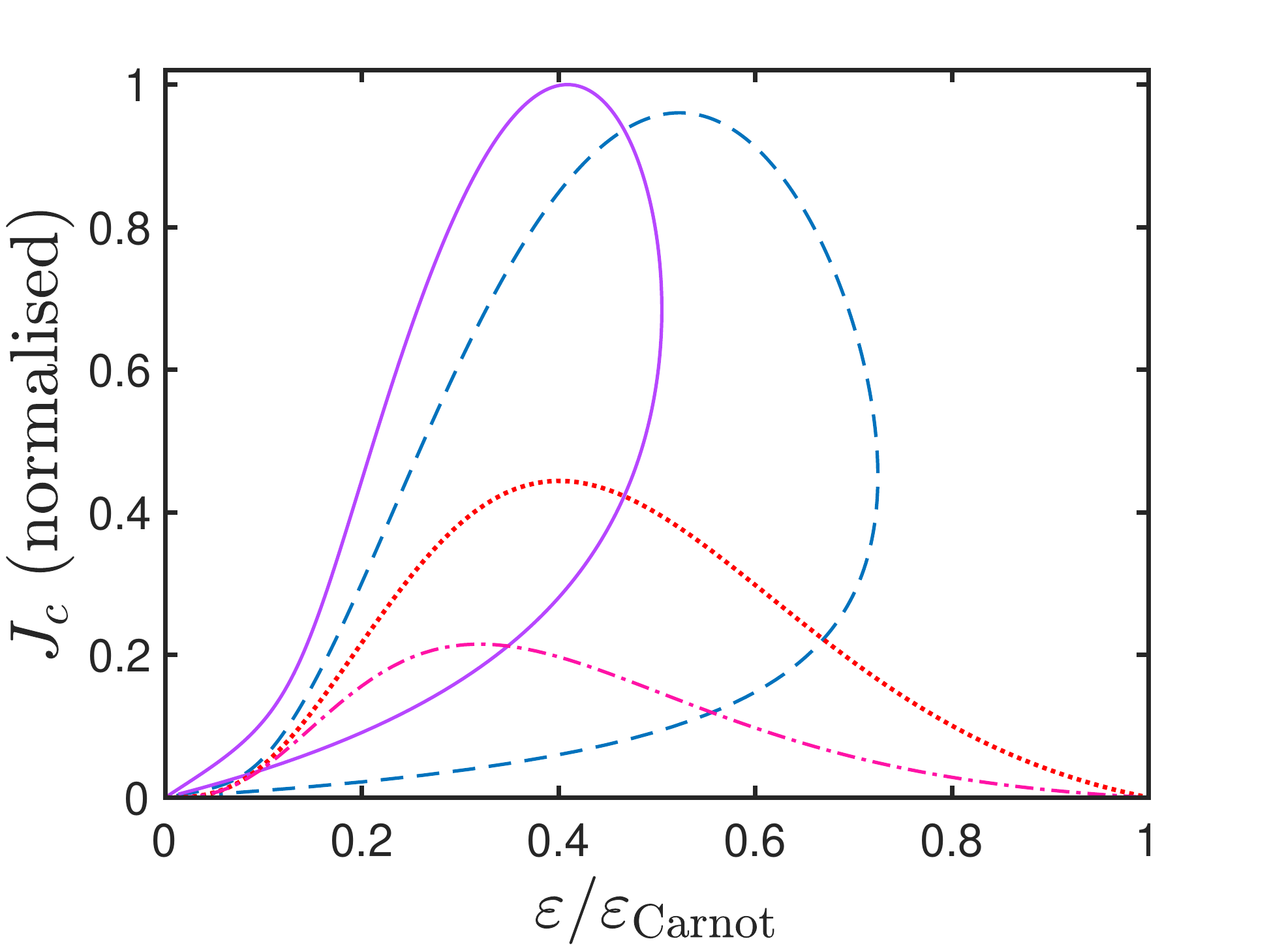}
	\end{minipage}
\begin{minipage}{0.45\linewidth}	
	\flushleft(b)\\
	\centering
	\includegraphics[width=\linewidth]{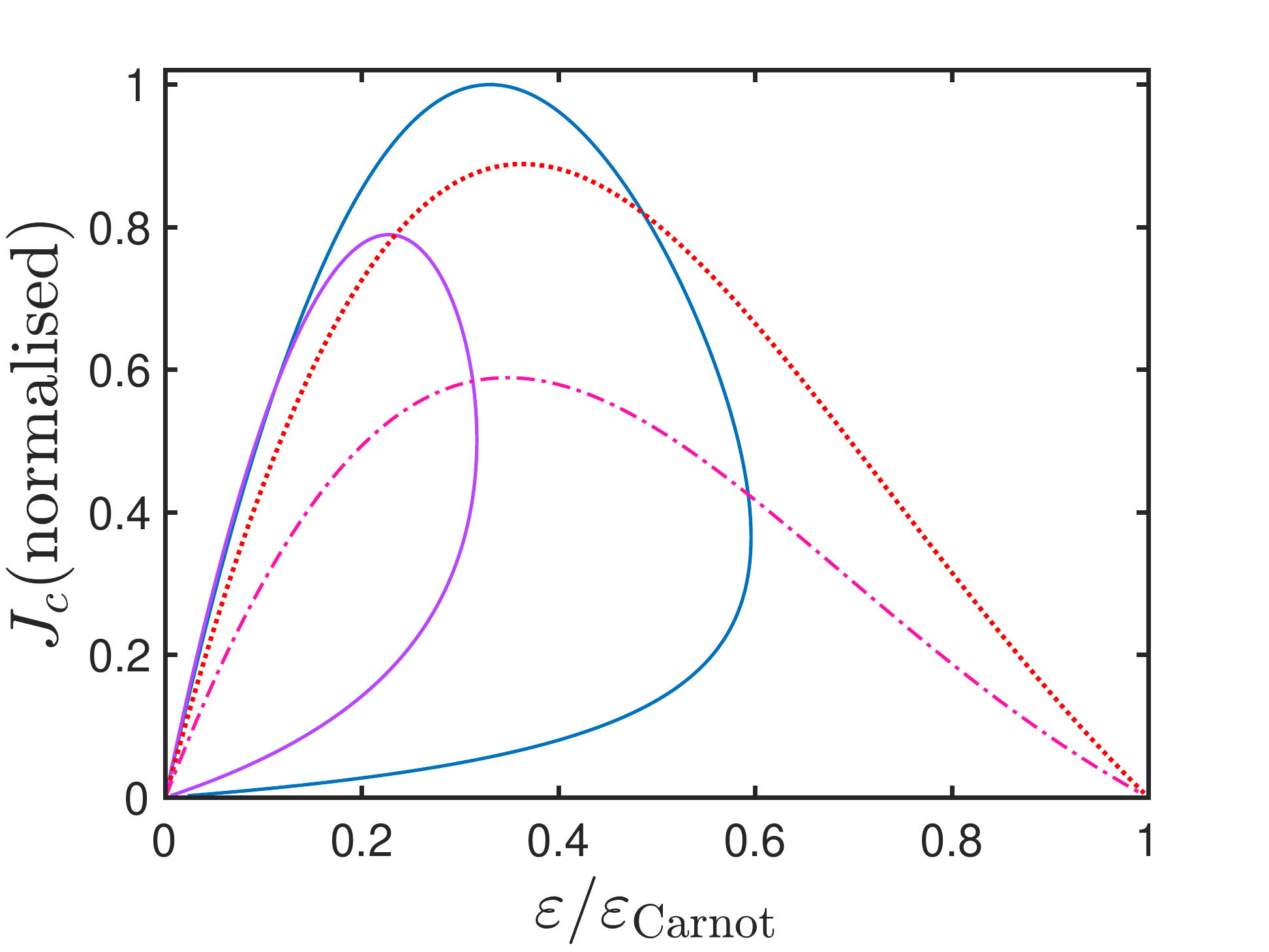}
\end{minipage}
	\caption{Typical performance characteristics of a three-qubit absorption refrigerator for $\{T_c,T_h,T_w\}=  \{1,1.1,1.5\}$, in units of $\omega_w$, for a $D$-dimensional environment with (a) $D=1$ and (b) $D=3$. The cooling power (normalised by the largest value) is plotted against coefficient of performance as the cold qubit frequency varies within the cooling window. The solid and dashed lines show $g = 0.4$ and $g= 0.2$, respectively, calculated with a global master equation. The dotted and dot-dashed lines show $g=0.04$ and $0.02$, respectively, computed with a local master equation. Details of the master equations are given in Ref.~\cite{Mitchison2015}. \label{fig:fridge_characteristic}}
\end{figure}

The trade-off between power and efficiency is illustrated by the typical characteristic curves plotted in Fig.~\ref{fig:fridge_characteristic}, which show power against efficiency as $\omega_c$ is varied within the cooling window~\eqref{cooling_window}. In general, weak-coupling refrigerators are able to achieve the highest efficiencies but at reduced cooling power. However, the functional dependence of power on coupling strength may be non-monotonic, in general. The open curves obtained from the local dissipation model are characteristic of endoreversible refrigerators, i.e. those without internal dissipation~\cite{Correa2015}. The closed curve calculated from the global approach is typical of realistic, i.e. non-ideal, thermal machines. 

One therefore expects on physical grounds that a similar closed curve should also be observed for the three-qubit refrigerator in the weak-coupling regime. Indeed, a key source of irreversibility not captured in Fig.~\ref{fig:fridge_characteristic} is broadening of energy levels induced by the coupling to the bath~\cite{Nakpathomkun2010,Josefsson2018}. This smears the sharp transition frequencies of the isolated system over a continuous range, whose width is proportional to the rates of dissipation. Since each frequency in this continuum sees a slightly different virtual temperature, perfectly reversible operation is again unachievable. However, level broadening is a higher-order effect that is not captured by standard perturbative master equation approaches. 

Since the Carnot point is generically unreachable and associated with zero power, a more appropriate figure of merit is the COP at maximum power, denoted $\varepsilon_*$. Remarkably, Correa and colleagues~\cite{Correa2014a,Correa2014b} were able to derive a bound on the COP at maximum power for a broad class of quantum absorption refrigerators, namely
\begin{equation}\label{COP_max_power_bound}
\frac{\varepsilon_*}{\varepsilon_{\rm Carnot}} \leq \frac{D}{D+1}.
\end{equation}
Here, $D$ is the spatial dimensionality of the bosonic environment, as discussed below Eq.~\eqref{detailed_balance_rates}. The bound~\eqref{COP_max_power_bound} applies in particular to the three-qubit refrigerator and all absorption refrigerators with endoreversible performance characteristics.

The aforementioned results illustrate stark differences between the predictions of the local and global master equations, which are only strictly valid in the extreme limits of strong or weak coupling. The regime of intermediate coupling was explored in Ref.~\cite{Seah2018}, using a master equation that interpolates between these two limits by varying a phenomenological coarse-graining time-scale~\cite{Schaller2008}. Focusing on the case $D=1$, it was shown that the optimal steady-state refrigerator performance is obtained at moderately strong coupling. This is because very weak coupling significantly reduces cooling power, while very strong coupling is associated with irreversible and off-resonant processes.

One may also ask whether alternative refrigerator designs can improve performance. Relaxing the condition of additive dissipation~\eqref{additive} allows for smaller refrigerators comprising two qubits~\cite{Levy2012,Levy2012a,Correa2014a,Silva2015} or even just one~\cite{Mu2017}. Here, the performance characteristics are similar to the three-level refrigerator. Alternatively, one can consider increasing the size of the machine by adding more energy levels. With an appropriate design, this leads to a reduction in the virtual temperature~\cite{Silva2016}. However, the cooling power can only be increased up to a saturation threshold (for fixed transition rates), while the efficiency is generally not improved~\cite{Correa2014c,Gonzalez2017}. A detailed mathematical analysis of general $N$-level absorption machines can be found in Ref.~\cite{Gonzalez2017}. For a comparison between various quantum absorption refrigerator designs and their power-driven counterparts, see Ref.~\cite{Clivaz2017}.

\subsection{Quantum performance enhancements?}
\label{sec:quantum_abs}

Quantum interference due to coherence between energy eigenstates is one of the principal differences between quantum thermodynamics and its classical counterpart. More generally, quantum mechanics is distinguished from classical physics by the possibility of strong correlations such as entanglement and discord. It is thus natural to ask whether these specifically quantum features might improve the performance of absorption refrigerators. 

Regarding quantum correlations, Brunner et al.~\cite{Brunner2014} studied the weak-coupling regime of the three-qubit refrigerator, finding that the presence of entanglement is associated with a higher cooling power but a smaller COP. Moreover, the entanglement vanishes when operating close to the Carnot point. Coherences have been shown to play a similar role, boosting power at the expense of efficiency, in two different models of a quantum absorption refrigerator that are closely related to the three-level and three-body refrigerators discussed above~\cite{Holubec2018,Du2018}. 

The basic intuition behind the results of Refs.~\cite{Brunner2014,Holubec2018,Du2018} is that energy currents traversing a multipartite system are associated with coherence in the energy eigenbasis~\cite{Mitchison2018}. (For example, the expectation value in Eq.~\eqref{coherence} is directly proportional to a coherence between eigenstates of the Hamiltonian~\eqref{Htot}.) Therefore, large amounts of coherence (which might also imply entanglement~\cite{Brunner2014}) are associated with more current and thus more cooling power, especially in the regime of weak coupling $g$ between parts of the machine. Note, however, that this conclusion is model-dependent: in certain configurations, decoherence can actually increase cooling power by suppressing destructive interference~\cite{Kilgour2018}, an effect that is well known in quantum biology~\cite{Huelga2013}.

At large $g$, on the other hand, excitations travel across the system more rapidly but the magnitude of the current and its associated coherence remains limited by the weak coupling to the reservoirs. (For example, consider increasing $g$ in Eq.~\eqref{coherence} while holding $k$ fixed.) Therefore, coherence in the strong-coupling regime is mainly associated with deleterious effects like destructive interference and internal dissipation, as discussed in the previous section and Refs.~\cite{Correa2013,Correa2015,Man2017,Kilgour2018}. In fact, coherence is associated quite generally with additional entropy production~\cite{Santos2019,Francica2019} and thus is not expected to offer improvements in efficiency.

Reservoir engineering offers a different route to seeing better performance by tailoring the quantum state of the baths. One such possibility is squeezing, in which the fluctuations of one quadrature (e.g. the momenta of mechanical modes in a vibrational bath) are suppressed while fluctuations in the other (e.g. the mode coordinates) increase. In particular, squeezing the work reservoir has been shown to enhance both the power and efficiency in quantum absorption refrigerators, compared to those operating between non-squeezed thermal baths at the same temperature~\cite{Correa2014a}. However, the preparation of a squeezed state requires additional work, which may not be used most efficiently in this way~\cite{Roulet2018a}. Reservoir engineering also demands significant control over the bath's microscopic degrees of freedom.

\begin{figure}
	\centering
	\includegraphics[trim=0mm 80mm 0mm 70mm, clip,width=0.8\linewidth]{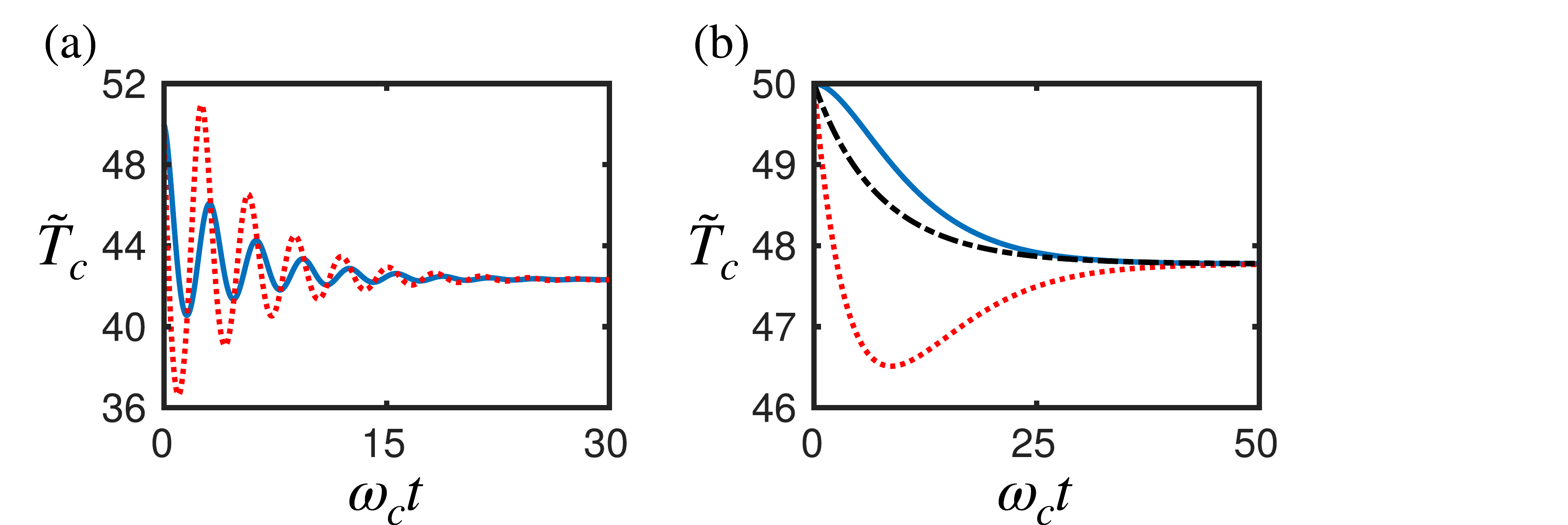}
	\caption{Temperature dynamics of the cold qubit in a three-qubit refrigerator in (a) the strong-coupling regime and (b) the weak-coupling regime. Solid lines show the evolution of an initial product state, where each qubit starts at equilibrium with its respective bath. Dotted lines show the dynamics with additional coherence in the initial three-qubit state. The dot-dashed line in (b) shows the equivalent stochastic refrigerator. Figure adapted from Ref.~\cite{Mitchison2015}.\label{fig:single_shot_cooling}}
\end{figure}

Another scenario in which coherence plays a role is single-shot cooling~\cite{Mitchison2015}, where the aim is to cool subsystem $c$ as quickly and effectively as possible before extracting it for use elsewhere. In three-body refrigerators, lower temperatures than the steady state can be reached in a finite time by harnessing coherent oscillations of the local energies~\cite{Mitchison2015,Brask2015}. As shown in Fig.~\ref{fig:single_shot_cooling}, this effect is boosted by adding more coherence to the initial state, demonstrating that coherence is a useful resource for some cooling tasks. Interestingly, in the weak-coupling regime of the three-qubit refrigerator, one can arrange for a slow, near-critically damped oscillation, which allows for a single-shot cooling advantage without precisely timed extraction of the cold subsystem~\cite{Brask2015}. In contrast, a classical stochastic refrigerator without any coherences in the energy eigenbasis, corresponding to the limit of very weak coupling and strong thermalisation, would only exhibit pure exponential relaxation, with the lowest achievable temperature in the asymptotic steady state~\cite{Mitchison2015,Nimmrichter2017}. Nevertheless, the temperature oscillations that enable single-shot cooling in a refrigerator comprising quantum harmonic oscillators can be reproduced in an analogous classical model of complex oscillators~\cite{Nimmrichter2017}. Similarly, one expects that single-shot cooling in the qubit refrigerator might also be possible in an analogous model of classical precessing spins.

The above considerations illustrate a general problem that arises when trying to identify quantum advantages in thermodynamics: there is no unique notion of the ``classical limit'' of a quantum thermal machine. This is quite different to, say, the field of computational complexity, where the concepts of a quantum and a classical computer are rigorously defined and their differences can be unambiguously analysed. Therefore, while coherent effects in a given microscopic process may be found to play an important role, e.g. because additional dephasing reduces power or efficiency~\cite{Uzdin2015,Kilgour2018}, it is usually possible to find another classical model that emulates the quantum machine by reproducing the same performance characteristics~\cite{Gonzalez2019}. We thus cannot yet decisively claim a \textit{generic} quantum advantage for thermal machines.

\subsection{Physical implementations}

Numerous experimental proposals have been put forward to realise the quantum absorption refrigerator across diverse platforms, including superconducting circuits~\cite{Chen2012,Hofer2016},  nanomechanical oscillators~\cite{Mari2012}, quantum dots~\cite{Venturelli2013,Erdman2018} and atom-cavity systems~\cite{Mitchison2016}. In each of these physical set-ups, it is possible to identify three subsystems that are coupled locally to independent heat baths and have an effective non-linear interaction allowing cooperative heat flow from the work and cold reservoirs into the hot one. 

\begin{figure}
	\centering
	\includegraphics[trim = 0mm 50mm 0mm 50mm, clip, width=0.7\linewidth]{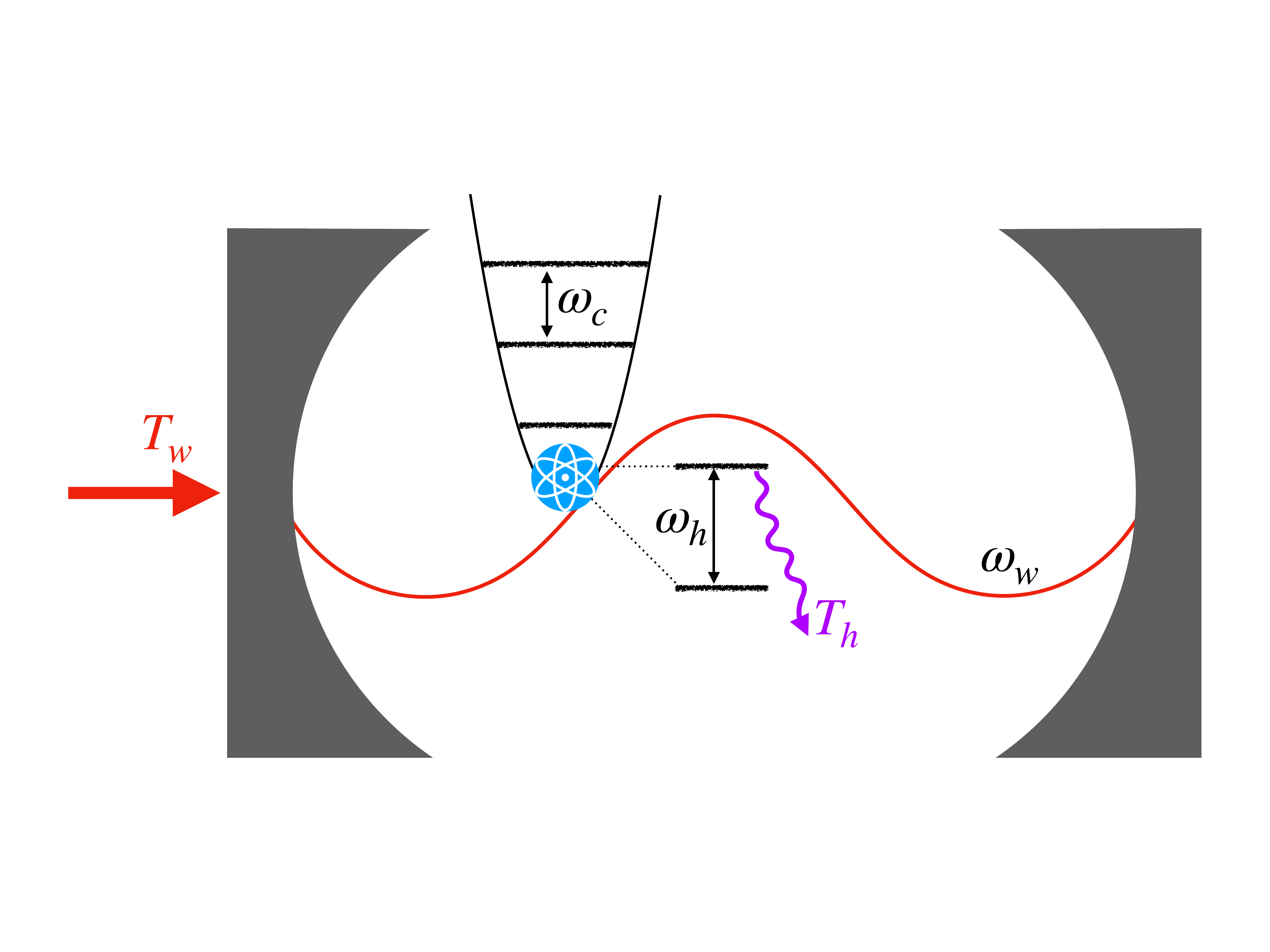}
	\caption{Schematic depiction of a quantum absorption refrigerator realised in cavity QED. A two-level atom is harmonically trapped near a node of the cavity field. The cavity field is maintained at high temperature by a thermal light beam. Cooling of atomic motion occurs because thermal cavity photons are absorbed and spontaneously re-emitted into the free radiation field at slightly higher frequency. \label{fig:atom_cavity}}
\end{figure}

The proposal of Ref.~\cite{Mitchison2016} is an example where such an interaction arises naturally. Consider a classic setting of cavity quantum electrodynamics (QED): a two-level atom confined by a harmonic potential inside an optical cavity~\cite{Buzek1997}, as illustrated in Fig.~\ref{fig:atom_cavity}. Let $\omega_c$ be the fundamental oscillation frequency of the atomic motion, $\omega_h$ be the electronic energy splitting and $\omega_w$ be the frequency of the cavity mode. Physically, a tripartite interaction between these degrees of freedom arises because photons carry both energy and momentum. Therefore, absorption or emission of a photon changes not only the atom's internal state but also its state of motion, in general. The effect of the cavity light on atomic motion is maximised when the trap potential minimum coincides with a node of the electric field. Choosing the cavity mode to be resonant with the red sideband frequency $\omega_w = \omega_h-\omega_c$ (c.f. Eq.~\eqref{resonanceCondition}), the dominant contribution to the light-matter interaction is then given by Eq.~\eqref{three_body} with  $\hat{A}_{c,w} = \hat{a}_{c,w}$ the annihilation operators for the atomic motion and the cavity mode and $\hat{A}_h = \hat{\sigma}^-_h$ the lowering operator for the two-level atom. This interaction drives transitions between the states
\begin{equation}\label{atom_cavity_transitions}
\ket{n}_c\ket{0}_h\ket{m}_w \longleftrightarrow \ket{n-1}_c \ket{1}_h \ket{m-1}_w.
\end{equation}

\begin{figure}
	\centering
	\includegraphics[trim = 0mm 40mm 0mm 40mm, clip, width=0.7\linewidth]{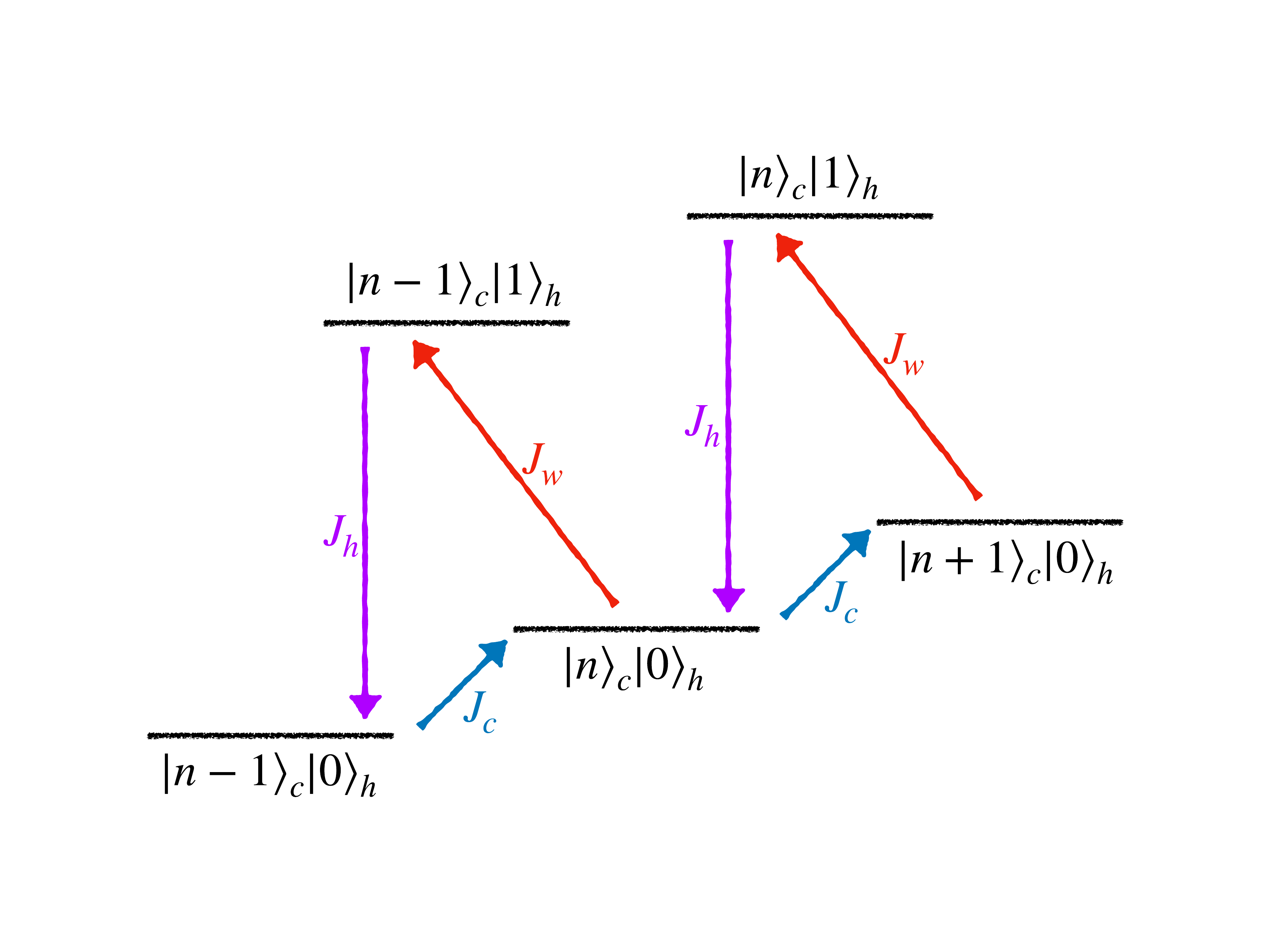}
	\caption{Scheme of electronic and vibrational levels of a trapped, two-level atom, illustrating the  cooling cycle of the cavity-QED refrigerator. Absorption of cavity-filtered photons from the work reservoir followed by emission into the free radiation field (i.e. the hot reservoir) leads to sideband cooling of the atomic motion.\label{fig:sideband}}
\end{figure}

Suppose now that the cavity is driven by a beam of thermal light at temperature $T_w$, populating the cavity with photons. The cavity behaves as a frequency filter that allows only those photons in the work reservoir that are resonant with the red sideband frequency $\omega_w$ to interact with the atom, which carry slightly less energy than $\omega_h$. Each time the atom absorbs a cavity photon, it must lose a quantum of vibrational energy to make up for the energy deficit. The electronic state is then reset by spontaneous emission of a photon into the free radiation field at temperature $T_h<T_w$. Meanwhile, the motion of the atom generally undergoes some heating with an effective temperature $T_c$. As shown in Fig.~\ref{fig:sideband}, so long as the heating rate is not too large the atomic motion is driven down towards the ground state. The cooling cycle is analogous to laser sideband cooling~\cite{Leibfried2003}, but with the laser field replaced by the thermal state of the cavity.

In this case, there are infinitely many virtual qubits, namely the pairs of states $\ket{0}_h\ket{m}_w$ and $\ket{1}_h\ket{m-1}_w$. Each of these pairs has the same energy splitting $\omega_c$ and virtual temperature given by Eq.~\eqref{virtual_temp}. The enormous difference between vibrational and optical frequencies makes very low virtual temperatures achievable. Indeed, let us make the natural assumption that $T_h = T_c = T_{\rm room} \approx 300~$K is the ambient temperature of the laboratory, and that $T_w\gg T_{\rm room}$. The virtual temperature is then approximately given by 
\begin{equation}\label{TvAtomCavity}
\frac{T_v}{T_{\rm room}} \approx \frac{\omega_c}{\omega_h}.
\end{equation}
For typical vibrational and optical frequencies, $\omega_c/\omega_h \lesssim 10^{-8}$, we therefore estimate $T_v~\sim~1~\mu$K. This points towards the exciting possibility of cooling a microscopic degree of freedom almost to its quantum ground state using only abundant sources of thermal energy: for example, focused sunlight at temperature $T_w\approx 5600$~K~\cite{Mitchison2016}.

Of course, many technical challenges must be overcome before this dream can be realised. In particular, active frequency stabilisation for the cavity is needed to maintain the relative difference $(\omega_h - \omega_c)/\omega_h = \omega_c/\omega_h$ to such a high degree of precision. Natural line broadening and motional heating due to spontaneous emission represent further practical limitations to the atom-cavity scheme, although these can be ameliorated with a more elaborate set-up~\cite{Mitchison2016}.

As for efficiency, since the atom-cavity refrigerator operates in the weak-coupling regime we may estimate the COP by the ideal value $\varepsilon = \omega_c/\omega_w$, which is of the same order as Eq.~\eqref{TvAtomCavity}, i.e. very small indeed! This tiny value merely expresses the fact that it is hugely wasteful, in energetic terms, to expend an optical photon in order to reduce the atom's motional energy by one quantum. This may be contrasted with other platforms where the subsystem energies are comparable to each other, such as the circuit-QED implementation proposed in Ref.~\cite{Hofer2016}, which can achieve greater efficiency but less fractional reduction in temperature (an explicit comparison can be found in Ref.~\cite{MitchisonPotts}). This illustrates a general trade-off for the quantum absorption refrigerator: cooling to very low temperatures can typically be achieved only at very low steady-state efficiencies, and vice versa.

The first experimental demonstration of a quantum absorption refrigerator was carried out in a trapped-ion system by Maslennikov and colleagues \cite{Maslennikov2019}. However, this experiment is somewhat different from the systems described above because there are no external baths.  Instead, the temperatures are defined by preparing the three subsystems in different thermal states at the start of the experiment. In particular, these subsystems are three vibrational modes of three trapped $^{171}$Yb$^+$ ions. The anharmonicity of the Coulomb interaction between the positively charged ions leads to coupling between these modes~\cite{Marquet2003}. Assuming that the resonance condition~\eqref{resonanceCondition} holds, the leading-order anharmonic perturbation is given by Eq.~\eqref{three_body}, with $\hat{A}_{c,h,w} = \hat{a}_{c,h,w}$ the usual harmonic oscillator annihilation operators. 

 After the initial preparation, the system is allowed to evolve unitarily. Remarkably, the reduced state of the $c$ mode nonetheless reaches a stationary condition, which is characterised by an effective temperature $T < T_c$. The appearance of local equilibrium states from global unitary dynamics is a rather generic phenomenon in quantum many-body systems that has been extensively studied in recent years~\cite{Polkovnikov2011,Eisert2015,Gogolin2016}. In this case, local equilibration is possible even in a few-body system of three ions due to destructive interference between many different oscillatory modes in their infinite-dimensional Hilbert space~\cite{Nimmrichter2017}. 

In addition to steady-state cooling, Maslennikov et al. were able to experimentally demonstrate the single-shot cooling advantage due to transient oscillations predicted in Refs.~\cite{Mitchison2015,Brask2015}. Moreover, the effect of squeezing the initial thermal state of the work mode was investigated~\cite{Correa2014a}. It was found that squeezing could force the system to transition from a regime in which mode $c$ was heated to one in which it was cooled. However, squeezed thermal states led to smaller reductions in temperature than thermal ones with the same initial energy, demonstrating once again that quantum resources do not necessarily lead to an advantage.

\section{Autonomous quantum heat engines}
\label{sec:engines}

\subsection{Work in autonomous quantum machines}
\label{sec:work}

A heat engine harnesses the flow of heat from a hot reservoir at temperature $T_h$ to a cold reservoir at temperature $T_c<T_h$ in order to perform work on a load. In the autonomous formulation of thermodynamics, this load is modelled explicitly as a dynamical system rather than as a mere parameter entering the quantum Hamiltonian. But how do we define the work done by one quantum system on another? 

From the point of view of macroscopic thermodynamics, work is defined as a transfer of energy between equilibrium systems without an associated change in thermodynamic entropy. Thus, a simple way to quantify work for non-equilibrium quantum systems is a transfer of energy that does not change the von Neumann entropy. In particular, energy changes associated with unitary transformations are identified entirely with work. However, the work load of an autonomous quantum engine does not evolve unitarily, in general, leading to a growth of entropy as well as energy. It is therefore necessary to use a more refined measure that explicitly distinguishes the useful work output from the entropic contribution. Here, we quantify work via the concept of \textit{ergotropy}~\cite{Allahverdyan2004}, which was originally defined in the context of unitary work extraction but has since proved extremely useful for describing non-unitary transformations as well~\cite{Binder2015}. 

Consider first the unitary case, in which the Hamiltonian of some well-controlled, closed quantum system is changed cyclically as a function of time, $\hat{H}(t)$, such that $\hat{H}(0) = \hat{H}(\tau) = \hat{H}$. This generates a unitary transformation $\hat{U}$ acting on the initial state $\rho$, i.e. $\rho \to \hat{U}\rho\hat{U}^\dagger$. Since this is an isentropic process, the difference in energy between the initial and final states is associated with the extraction of work by the external control system:
\begin{equation}\label{W_unitary}
W_{\hat{U}} = \Tr \left \lbrace \hat{H} \left (\rho - \hat{U}\rho\hat{U}^\dagger \right ) \right \rbrace.
\end{equation}

A state is called ``passive'' with respect to the Hamiltonian $\hat{H}$ if no work can be extracted from it, i.e. if $W_{\hat{U}} \leq 0$ for any $\hat{U}$. All passive states have the form~\cite{Pusz1978}
\begin{equation}\label{passive_state}
\pi = \sum_k p_k \ketbra{E_k}{E_k},
\end{equation}
where $\ket{E_k}$ is an eigenvector of $\hat{H}$ with eigenvalue $E_k$, and the probabilities and energies are ordered as $p_k> p_{k+1}$ and $E_k < E_{k+1}$. In other words, the ground state has the highest population, the first excited state has the next-highest population, and so on. For example, a thermal equilibrium state at any (positive) temperature is passive.\footnote{Indeed, equilibrium states span the set of completely passive states, i.e. those which remain passive under composition, such that $\rho^{\otimes n}$ is passive for all $n$~\cite{Pusz1978,Lenard1978,Skrzypczyk2015}.} This is of course consistent with Thomson's formulation of the second law, which forbids cyclic work extraction from thermal equilibrium states~\cite{Allahverdyan2002}.

The ergotropy $W$ is defined as the maximum work that can be extracted by a cyclic unitary, namely $W = \max_{\hat{U}} W_{\hat{U}}$. The maximum is clearly obtained when the final state is passive, i.e. $\hat{U}\rho\hat{U}^\dagger = \pi(\rho)$, where $\pi(\rho)$ is the unique passive state with the same eigenvalue spectrum as $\rho$.  In particular, writing $\rho = \sum_k p_k \ketbra{\psi_k}{\psi_k}$ and $\hat{H}=\sum_k E_k\ketbra{E_k}{E_k}$, we have the explicit expression for the ergotropy
\begin{align}\label{ergotropy}
W(\rho,\hat{H}) & = \Tr\left \{\hat{H}\left [\rho - \pi(\rho)\right ]\right \}\notag\\
&  = \sum_{k,l} p_k E_l \left ( |\braket{E_l}{\psi_k}|^2 - \delta_{kl}\right ),
\end{align}
where $p_k> p_{k+1}$ and $E_k < E_{k+1}$ as before. We see that ergotropy quantifies the energy content associated with population inversion or coherence in the energy eigenbasis.

Now, the evolution of a load connected to an absorption engine is \textit{not} unitary, in general, leading to the production of entropy. As a consequence, not all of the energy transferred to the load can be identified with work. Instead, we quantify work as the change in the load's ergotropy. In particular, if $\rho_w$ is the state of the work load with Hamiltonian $\hat{H}_w$, which is transformed by the engine into a new state $\rho_w'$, the work done is
\begin{equation}\label{work_done}
\Delta W =  W(\rho_w',\hat{H}_w) -W(\rho_w,\hat{H}_w).
\end{equation}
It follows that an absorption machine performs work on the load by boosting the population of high-energy states relative to low-energy ones or by generating coherence in the energy eigenbasis. On the other hand, purely entropic processes such as cooling or heating, which transfer the load between passive states, do not constitute work. Eq.~\eqref{work_done} thus quantifies an \textit{ordered energy increase}, as we shall see in the following section. 

However, it is important to point out that work does not have a uniquely agreed-upon definition in the quantum domain. While the present definition appears to be a useful one for describing autonomous quantum engines~\cite{Gelbwaser-Klimovsky2014,Gelbwaser-Klimovsky2015,Seah2018a,Loerch2018}, other possibilities have been considered in this setting~\cite{Boukobza2006a,Weimer2008,Brunner2012,Skrzypczyk2014,Mari2015}. Work can even be defined more generally to incorporate the fluctuations between individual experimental realisations (see, for example, Refs.~\cite{Talkner2007,Dahlsten2011,Horodecki2013,Aberg2013,Egloff2015,Halpern2015,Guarnieri2019a}). We also emphasise that, although the definition of ergotropy~\eqref{ergotropy} makes reference to a non-autonomous unitary process, Eq.~\eqref{work_done} applies to an arbitrary transformation of a quantum system with a fixed Hamiltonian. Nevertheless, it is admittedly unsatisfying that in order to define work production in autonomous machines we have appealed to the (fictitious) non-autonomous work-extraction process described by Eq.~\eqref{W_unitary}. Formulating work extraction from quantum states in a general, fully autonomous framework is an important foundational problem for quantum thermodynamics~\cite{Aberg2018}. 

\subsection{Work production in absorption engines}
\label{sec:three_body_eng}

\begin{figure}
	\centering
	\includegraphics[trim = 80mm 80mm 80mm 80mm, clip,width=0.4\linewidth]{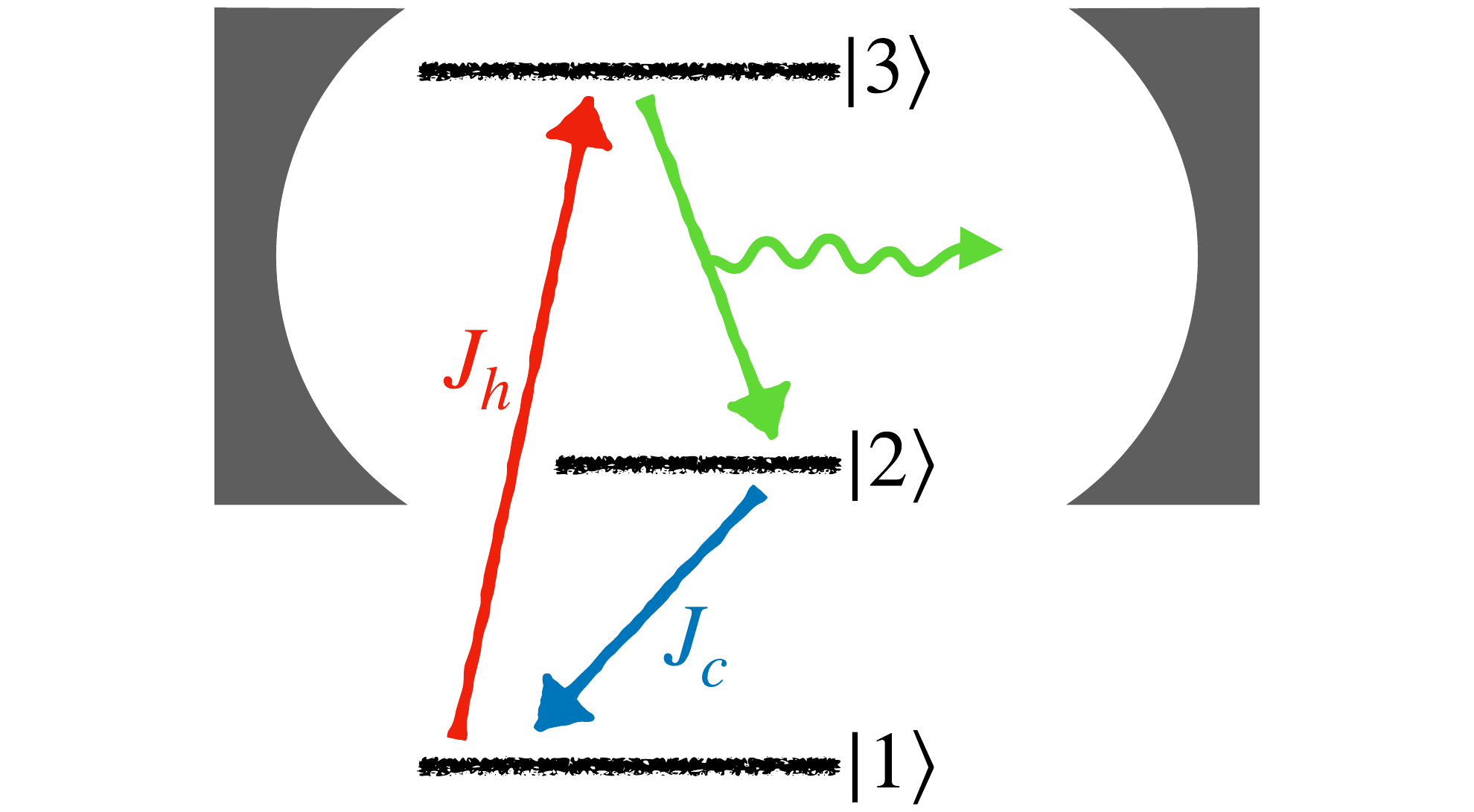}
	\caption{A three-level absorption engine coupled to a cavity load. The flow of heat from a hot to a cold bath is partially converted into work by the emission of photons into the cavity. \label{sec:three_level_cavity}}
\end{figure}

A straightforward way to design an autonomous quantum heat engine is to run a quantum absorption refrigerator in reverse. Arguably, the simplest such engine involves a three-level system interacting with filtered thermal reservoirs and a load comprising a quantised cavity field~\cite{Boukobza2006,Ansari2017,Perl2017,Yuge2017}. This is a basic model of a laser, which generalises the Scovil-Schulz-DuBois model~\cite{Scovil1959} to incorporate the load explicitly.  The operation of the machine can be understood intuitively from Fig.~\ref{sec:three_level_cavity}. The flow of heat from the hot bath to the cold one generates a population inversion on the work transition. This leads to a sustained stimulated emission of photons into the cavity. However, the dissipative forcing from the thermal baths also generates a continuous growth of entropy in the state of the cavity field~\cite{Boukobza2006}. This implies that some of the energy transferred to the cavity should properly be considered as heat, rather than work.

To elucidate how heating in the load affects work output, we focus on an alternative set-up that explicitly incorporates the filtering of the baths. This necessitates two distinct subsystems $c$ and $h$ for the cold and hot baths, respectively, in addition to a load, $w$. Each subsystem is characterised by a single transition frequency $\omega_{c,h,w}$ and has local Hamiltonian given by Eq.~\eqref{Halpha}. The frequencies are assumed to satisfy the resonance condition~\eqref{resonanceCondition}, i.e. $\omega_w = \omega_h - \omega_c.$ The interaction Hamiltonian describing the absorption engine reads as 
\begin{align}
\label{three_body_eng}
\hat{H}_{\rm int} & = g \left (\hat{A}^\dagger_c \hat{A}_h  \hat{A}^\dagger_w  + \hat{A}^\dagger_c \hat{A}_h \hat{A}^\dagger_w \right ),
\end{align}
where $\hat{A}_\alpha$ is a lowering operator satisfying $[\hat{H}_\alpha,\hat{A}_\alpha] = -\omega_\alpha \hat{A}_\alpha$. This Hamiltonian allows the resonant transfer of heat from $h$ to $c$, but only by increasing the energy of $w$ (c.f.~Eq.~\eqref{three_body}). Such a model was first studied by Youssef and colleagues~\cite{Youssef2009,Youssef2010}, with numerous variants subsequently introduced~\cite{Brunner2012,Mari2015,Levy2016}.

\begin{figure}
\centering
\includegraphics[trim = 50mm 80mm 50mm 40mm, clip,width=0.5\linewidth]{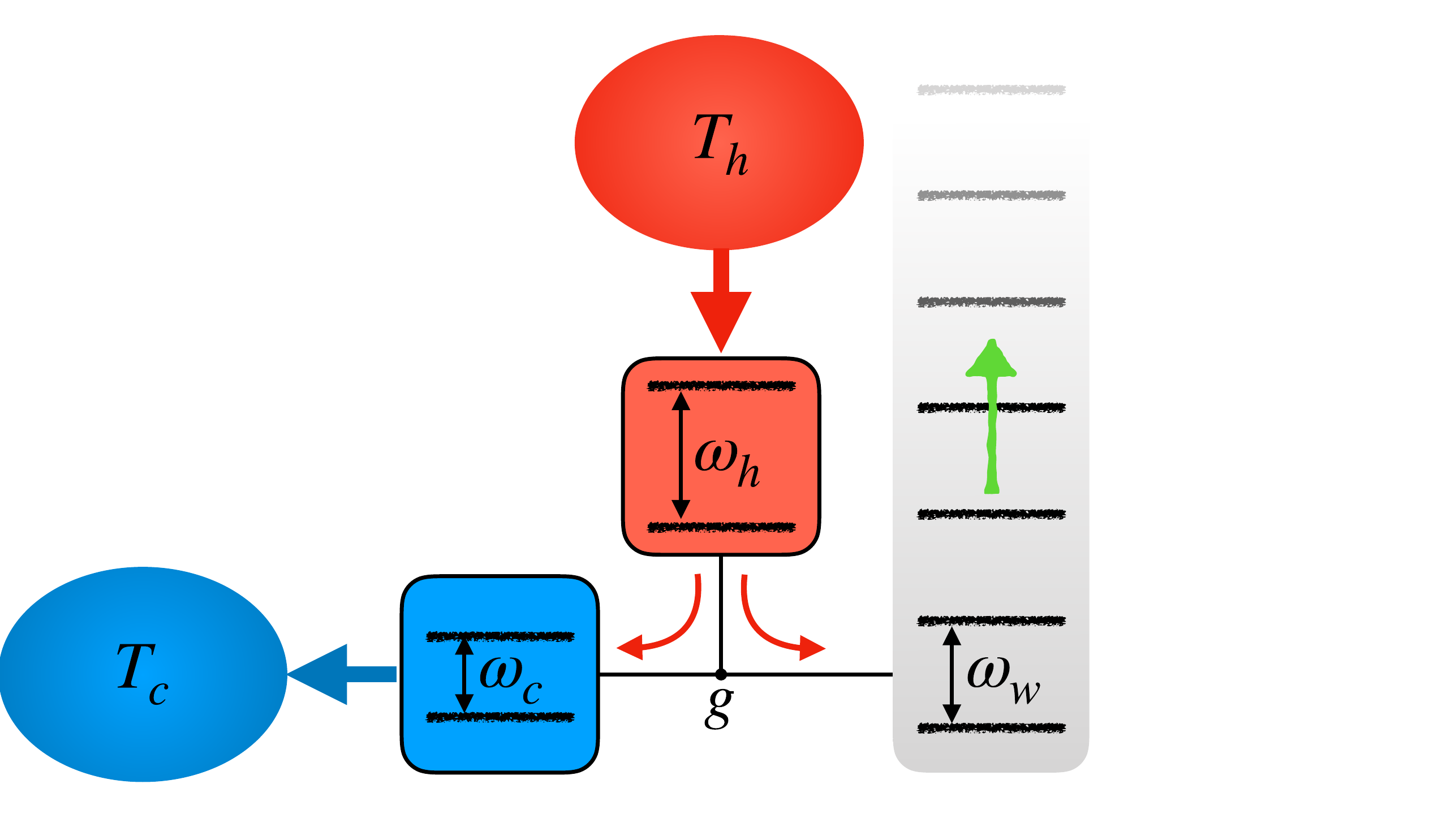}
\caption{A two-qubit absorption engine captures a portion of the heat flowing from a hot bath to a cold one in order to perform work on a load,  modelled here as an infinite ladder.  \label{fig:heat_engine}}
\end{figure}

\begin{figure}[b]
	\centering
	\includegraphics[trim = 70mm 50mm 70mm 50mm, clip,width=0.4\linewidth]{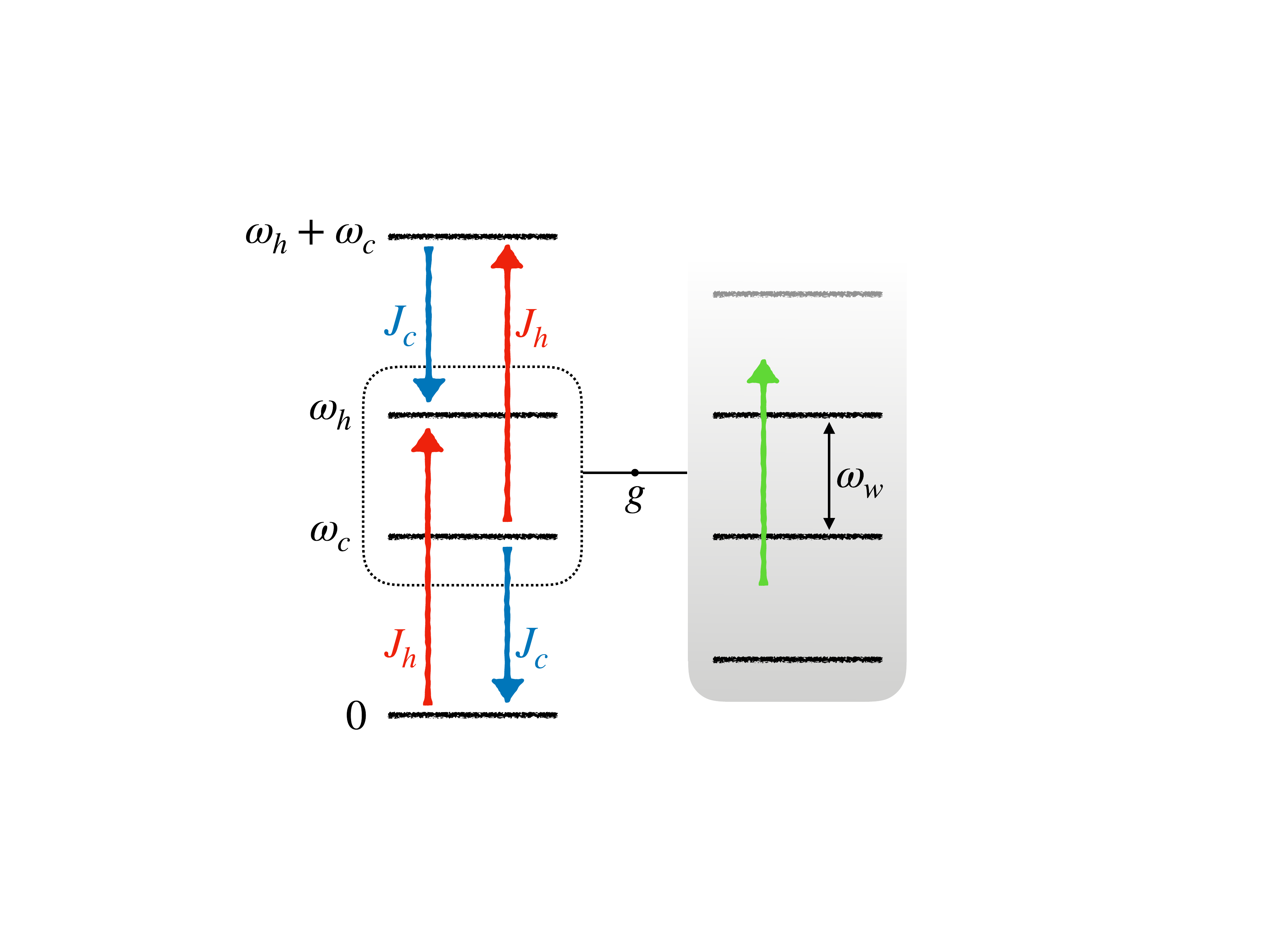}
	\caption{In an engine, the heat currents induce population inversion in the virtual qubit, thus driving the load up the ladder.\label{fig:virtual_qubit_eng}}
\end{figure}

For simplicity, we take $c$ and $h$ to be two-level systems, while the load is modelled as an infinite ladder of equally spaced states, as illustrated in Fig.~\ref{fig:heat_engine}. In particular, it is convenient to assume that the ladder operator for the load is $\hat{A}_w= \sum_{n=0}^\infty \ket{n}_w\bra{n+1}$, whose matrix elements are independent of $n$. The interaction Hamiltonian drives transitions such as 
\begin{align}\label{engine_transitions}
\ket{0}_c\ket{1}_h\ket{n}_w \longleftrightarrow \ket{1}_c\ket{0}_h\ket{n+1}_w.
\end{align}
This shows that the load interacts with a two-dimensional subspace of states $\{\ket{0}_c\ket{1}_h, \ket{1}_c\ket{0}_h\}$ constituting the machine's virtual qubit. The virtual qubit's transition frequency is $\omega_h - \omega_c = \omega_w$ so that it exchanges energy resonantly with the load. When $c$ and $h$ are close to equilibrium with their respective baths, the virtual qubit states are populated in the ratio
\begin{equation}\label{virtual_ratio_eng}
\frac{p^{(0)}_c p^{(1)}_h}{p^{(1)}_c p^{(0)}_h} = \ee^{-\beta_v \omega_w},
\end{equation}
where $\beta_v = 1/T_v$ is the inverse of the engine's virtual temperature
\begin{equation}\label{virtual_temp_eng}
T_v = \frac{\omega_h - \omega_c}{\beta_h \omega_h - \beta_c \omega_c}.
\end{equation}
Switching on a weak coupling to the load, the machine induces population inversion and thus performs work so long as $T_v<0$. Indeed, a negative virtual temperature ensures that the population of $\ket{0}_c\ket{1}_h$ is larger than that of $\ket{1}_c\ket{0}_h$, which makes upwards transitions in the ladder more probable, as depicted in Fig.~\ref{fig:virtual_qubit_eng}. This is analogous to a classical heat engine lifting a weight. Here, however, the system is small enough that thermal fluctuations can introduce significant uncertainty in the state of the load (the position of the weight), which generally reduces work output as we now discuss. 

Let us assume that the load is initialised in a low-energy passive state, such as its ground state or a thermal state at one of the bath temperatures. The machine does not induce any local coherences in the load, so that its state has the diagonal form $\rho_w(t) = \sum_n p_n(t) \ket{n}_w\!\bra{n}$. A simple picture of the load dynamics is obtained in the limit where the qubits thermalise rapidly in comparison to the rate of energy transfer $g$ to the load. In this case, the fast qubit dynamics can be adiabatically eliminated~\cite{Erker2017} to obtain an effective equation of motion for the load's populations, which reads as
\begin{equation}\label{biased_random_walk}
\dot{p}_n = \gamma_\downarrow ( p_{n+1}- p_n) + \gamma_\uparrow\left (p_{n-1}- p_n\right ).
\end{equation}
This describes the probability distribution of a random walk on the ladder, where $\gamma_\uparrow$ and $\gamma_\downarrow$ denote the rate at which steps occur upwards and downwards, respectively. These rates depend on the temperatures and energy scales of the machine in such a way that $\gamma_\uparrow/\gamma_\downarrow = \ee^{-\beta_v\omega_w}$. Therefore, a net upwards motion is guaranteed if the virtual temperature is negative. 

Consider the limit of long times, so that the effect of the boundary at $n=0$ can be neglected. It follows from the central limit theorem that the occupation probabilities $p_n$ may be asymptotically approximated by a Gaussian distribution~\cite{vanKampen} with mean and variance
\begin{align}
\label{gaussian_mean}
\bar{n}(t) &= (\gamma_\uparrow - \gamma_\downarrow)t, \\
\label{gaussian_var}
\sigma^2(t) & = (\gamma_\uparrow + \gamma_\downarrow)t,
\end{align}
valid for times large enough so that $\bar{n}(t) \gg \sigma(t)$ and that memory of the initial condition is lost. It follows immediately that the mean and standard deviation of the load's energy are given by $E_w(t) = \omega_w \bar{n}(t)$ and $\Delta E_w(t) = \omega_w\sigma(t)$~\cite{Brunner2012}. The square-root growth of fluctuations in time, characteristic of a diffusion process, indicates that the entropy of the load is increasing (in fact, $S[\rho_w] \sim \log[(\gamma_\uparrow +\gamma_\downarrow)t]$). 

In the weak-coupling regime, each quantum of energy transferred to the load corresponds to one quantum absorbed from the hot bath. Therefore, the rate of energy increase in the load $\dot{E}_w$ is related to the input heat current $J_h$ by the ratio
\begin{equation}\label{ratio}
R = \frac{\dot{E}_w}{J_h} = 1 - \frac{\omega_c}{\omega_h}.
\end{equation}
The fact that $T_v<0$ implies that this ratio is bounded by the Carnot efficiency
\begin{equation}\label{Carnot_bound_eng}
R < \eta_{\rm Carnot} = 1 - \frac{T_c}{T_h}.
\end{equation}
However, the efficiency is smaller than this ratio because a portion of the energy passed from the hot bath to the load constitutes heat.

The actual power transferred to the load is the time derivative of its ergotropy $W(\rho_w,\hat{H}_w)$. In order to compute this, we need the corresponding passive state at each moment in time, $\pi(\rho_w) = \sum_n q_n \ket{n}_w\!\bra{n}$. For simplicity, let us focus on those instants where $\bar{n}=N$ is an integer. This is equivalent to considering a discretised time scale $t_N = N\Delta t$, where $N$ is the number of ``cycles'' of duration $\Delta t = (\gamma_\uparrow - \gamma_\downarrow)^{-1}$, during which the load's mean energy increases by $\omega_w$. At these instants in time, the passive state's occupation probabilities are given by $q_{2m} = q_{2m+1} = \left (2\pi \sigma^2\right )^{-1/2}\ee^{-m^2/2\sigma^2}$ for $m = 0,1,2,\ldots$. The corresponding ergotropy is found to be $W_N \approx E_w(t_N) - 4\Delta E_w(t_N)/\sqrt{2\pi}$, up to order $N^{1/2}$. The power transferred to the load on this coarse-grained time scale is given by $\dot{W}_N = (W_N - W_{N-1})/\Delta t$. We thus obtain the asymptotic result for the $N$-cycle efficiency $\eta_N = \dot{W}_N/J_h$, 
\begin{align}\label{efficiency_engine}
\eta_N  = 
\left [1 -  \sqrt{\frac{2\coth(-\beta_v\omega_w/2)} { \pi N}} \,  \right ]R.
\end{align}
This efficiency tends to the ideal ratio $R$ in the limit $N\to \infty$, where the energy transferred to the load $E_w\sim N$ is infinite. For any finite $N$, the efficiency is reduced by a correction of order $N^{-1/2}$, which becomes increasingly relevant as the Carnot point $T_v\to 0$ is approached.

\begin{figure}
	\centering
	\includegraphics[ clip, width=0.45\linewidth]{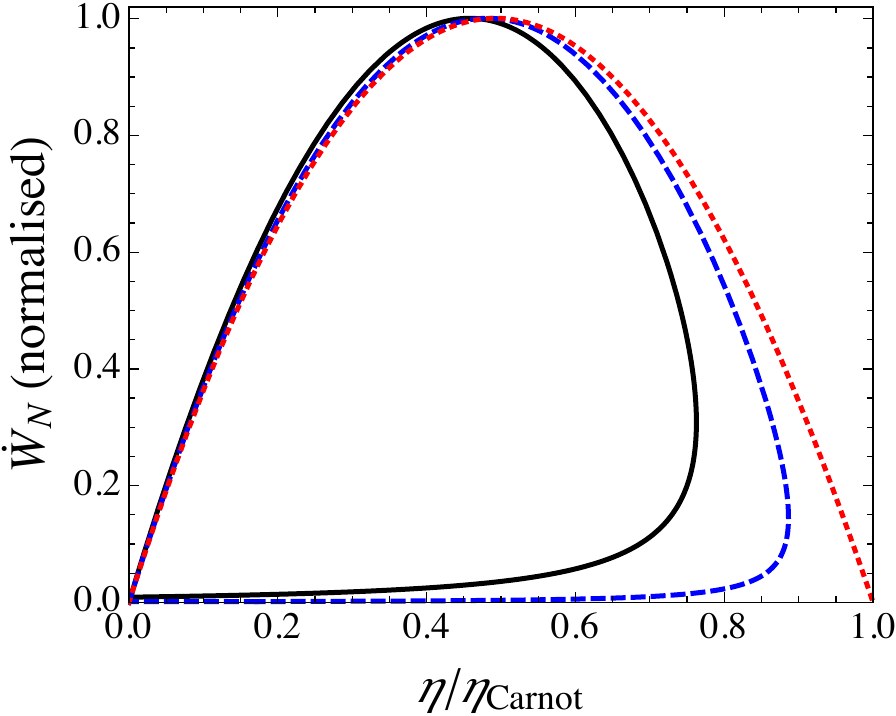}
	\caption{$N$-cycle performance characteristic of the two-qubit absorption engine in the weak-coupling limit, showing ergotropic power versus efficiency for $T_c = 1$ and $T_h = 10$, in units of $\omega_h$, for $N=300$ (solid line), $N=3000$ (dashed line) and $N\to\infty$ (dotted line). The curves are computed by varying $\omega_w$ within the regime of engine operation and using the random-walk model of the load dynamics detailed in Ref.~\cite{Erker2017}.\label{fig:engine_performance}} 
\end{figure}

The $N$-cycle performance characteristic of the engine is plotted in Fig.~\ref{fig:engine_performance}. For any finite number of cycles, heating of the load obstructs the Carnot efficiency from being reached. Only as $N\to \infty$ do we recover the open characteristic curve typical of an endoreversible machine. This shows that thermal fluctuations have a significant effect on work extraction unless a very large amount of energy is already invested. It is important to note that here, unlike for refrigeration, we see internal dissipation even at extremely weak coupling due to internal heating in the load. At stronger coupling, all of the coherent and irreversible effects discussed in Sec.~\ref{sec:fridge_performance} apply here as well because the engine is just an absorption refrigerator run in reverse. 

The strong similarity with absorption refrigerators means that the majority of experimental proposals mentioned in Sec.~\ref{sec:quantum_abs} could be straightforwardly adapted to work as an engine by arranging for a negative virtual temperature. An explicit physical implementation of an absorption engine was put forward in Ref.~\cite{Mari2015}, in the context of cavity optomechanics. Heat engines with a quantum load have been experimentally demonstrated using trapped ions, albeit with interactions controlled externally rather than autonomously~\cite{Lindenfels2018,Horne2018}. These experiments were nonetheless able to demonstrate the role of heating within the load, and represent an important step towards the realisation of a fully autonomous quantum heat engine. 

\section{Autonomous quantum clocks}
\label{sec:clocks}

\subsection{The cost of good timekeeping}
\label{sec:cost_timekeeping}

Time is one of the most fascinating and perennially surprising concepts in all of natural philosophy. Breakthroughs in our knowledge of the Universe, such as relativity theory, have been connected to changes in our understanding of time. Similarly, an improved capability to measure time has historically been associated with technological and economic progress. In the modern era, for example, the unprecedented precision of atomic clocks, which itself derives from fundamental energy quantisation, underpins satellite navigation and high-speed financial transactions as well as accurate scientific measurement standards. 

It is therefore important to determine whether there are any fundamental costs associated with accurate and precise timekeeping. This problem is not traditionally considered in the context of macroscopic thermodynamics, most likely because the power consumption of, say, a station clock is negligible compared to that of the trains whose comings and goings the clock dictates. In contrast, questions regarding the cost of timekeeping arise naturally in nanoscale thermodynamics, where even tiny amounts of energy can have a significant impact on efficiency. In the quantum case, moreover, the clock necessarily becomes correlated with the rest of the system, which may have a deleterious effect on its performance. 

Some examples of quantum clocks used to control heat engines are discussed in Refs.~\cite{Malabarba2015, Frenzel2016, Woods2018}. Here, the Hilbert-space dimension of the clock emerges as a limiting factor on performance, with perfect operation attainable only in the limit of infinite dimension and infinite energy~\cite{Woods2018}. Also notable in this regard are the ``pistons'' and ``rotors'' used to regulate quantum reciprocating engines~\cite{Tonner2005,Gelbwaser-Klimovsky2013,Gelbwaser-Klimovsky2014,Gelbwaser-Klimovsky2015,Mari2015,Roulet2017,Seah2018a,Roulet2018}.

More generally, various toy models of clocks have been proposed with the aim of exploring fundamental limitations imposed by quantum mechanics~\cite{Salecker1958,Peres1980,Woods2018}. Perhaps the most basic example consists of a single precessing spin, which was suggested to measure the duration of a quantum tunnelling process~\cite{Baz1967,Hauge1989}. In such models, an effective time scale arises from the evolution of a non-equilibrium state under a time-independent global Hamiltonian. Quantum mechanics dictates speed limits on the rate of such a clock's evolution in terms of the mean and variance of its energy~\cite{Mandelstam1991,Margolus1998}, while Lieb-Robinson bounds~\cite{Lieb1972} and, ultimately, Lorentz invariance restrict the speed of signal propagation. However, the aforementioned systems are typically designed only to measure a time interval and are thus more akin to a stopwatch than a clock.

\begin{figure}
	\centering
	\includegraphics[width=0.7\linewidth]{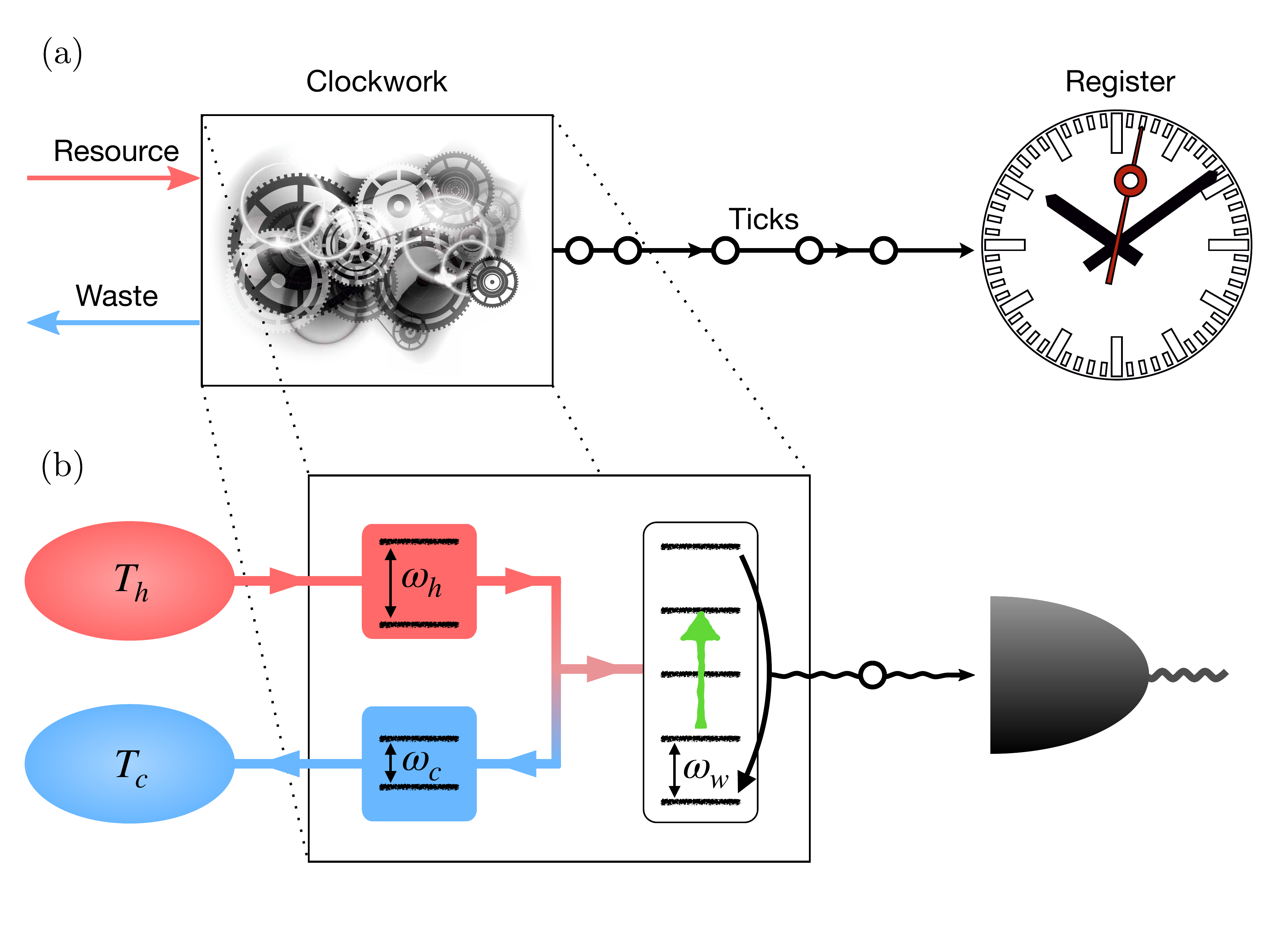}
	\caption{(a) A quantum clock comprises some physical clockwork that generates ticks, which are recorded by a register. A complete model of an autonomous clock must specify the resources needed to maintain the production of ticks. (b) The specific autonomous clock model proposed in Ref.~\cite{Erker2017}, involving a two-qubit heat engine driving a finite-dimensional ladder. The topmost level of the ladder is unstable and decays back to the ground state, emitting a photon that is recorded as a physical tick of the clock. Figure adapted from Ref.~\cite{Erker2017}. \label{fig:clock}}
\end{figure}

Indeed, an essential characteristic of a clock is that it continuously generates a time reference for an external observer. Since observation implies disturbance in quantum mechanics, a complete description of a quantum clock should explicitly specify the mechanism of readout and its back-action on the clock~\cite{Salecker1958}. This leads us to consider a clock as a bipartite system, comprising some internal ``clockwork'' whose dynamics generates a sequence of ticks, which are recorded by a ``register''~\cite{Rankovic2015}, as illustrated in Fig.~\ref{fig:clock}. In order for the clock to function autonomously, the transfer of information between clockwork and register should be irreversible and spontaneous (i.e. without the need for external intervention). This naturally suggests a connection to the second law of thermodynamics, since irreversible processes are associated with the production of entropy. 

Previous work has suggested that each tick of a quantum clock must generate entropy~\cite{Sels2015} and that the precise synchronisation of two quantum clocks is associated with entropy production~\cite{Janzing2003}. However, in order to obtain a complete description of a quantum clock one must specify the resources needed to initialise and maintain it. With this goal in mind, Erker and colleagues~\cite{Erker2017} introduced a fully self-contained model of a quantum clock and showed that its entropy production is a limiting factor on performance. That is, the more accurate and precise the clock is, the more energy it must expend and the more entropy it must produce per tick. The Hilbert-space dimension of the clock is also a key resource, with larger clocks able to achieve higher accuracy.

The above conclusions follow from a basic question: what are the \textit{minimal} resources needed to maintain an autonomous quantum clock? In order to ensure fair bookkeeping, these resources should not themselves require a clock to be prepared. In particular, the clock should evolve according to a time-independent Hamiltonian and the clock's power supply should work without any precise timing or control. This naturally points to thermal baths as the minimal resource to power the clock. Indeed, thermal states can be prepared deterministically without any timed control or understanding of the bath's microscopic dynamics. All other resource states have lower entropy for the same energy and thus require more knowledge to prepare. Since a clock is necessarily out of equilibrium, at least two baths at different temperatures are necessary. The heat flow between them can then be used to power the clock. In other words, the minimal self-contained quantum clock is a thermal absorption machine.

\subsection{Thermal absorption clock}

We now discuss the model of an autonomous quantum clock introduced in Ref.~\cite{Erker2017}. The machine is a modification of the heat engine discussed in Sec.~\ref{sec:three_body_eng}. Specifically, the clockwork comprises a two-qubit heat engine that performs work on a finite-dimensional load. Qubits $c$ and $h$ have frequencies $\omega_c$ and $\omega_h$ and are coupled to heat baths at temperatures $T_c$ and $T_h$, respectively. The load is modelled as a ladder of $d$ states, with adjacent levels separated in energy by $\omega_w = \omega_h-\omega_c$. As discussed in Sec.~\ref{sec:three_body_eng}, the engine induces population inversion in the load so long as $T_v <0$, where the virtual temperature is given by Eq.~\eqref{virtual_temp_eng}. 

In order to couple the clockwork to a register, we assume that the uppermost level of the ladder is unstable. Once the load reaches the top of the ladder it is able to decay back to the ground state by emitting a photon of frequency $(d-1)\omega_w$. We assume that $(d-1)\omega_w\gg T_h$ so that the reverse process of photon absorption from the environment is negligibly improbable. This results in a sustained stream of photons emitted by the clockwork. The register is a photodetector, which is assumed to detect the emitted photons with unit efficiency. Each detected photon corresponds to a single tick of the clock. 

As discussed in Sec.~\ref{sec:three_body_eng}, the evolution of the load is not deterministic, which leads to a stochastic sequence of ticks. The statistical properties of this sequence determine the performance of the clock. In particular, the clock's resolution gives the rate at which it ticks, while the clock's accuracy is, roughly speaking, the number of reliable ticks that it can produce. Note that here we wish to examine the emergence of a time scale from timeless (i.e. equilibrium) resources. We therefore quantify the the clock's accuracy relative to its \textit{own} ticks and not in absolute terms, in order to minimise the dependence on some assumed perfect background time scale. A rigorous approach to quantifying resolution and accuracy that completely avoids reference to background time would involve comparing the ticks produced by two initially synchronised clocks~\cite{Rankovic2015}, however it is difficult to perform explicit calculations in that framework (see, however, Ref.~\cite{Stupar2018}).

Our analysis is simplified by assuming that the clockwork is reset to precisely the same state after each photon emission. The photon detection times, $t_1,t_2,\ldots$, are then a renewal process, which means that the waiting times $w_n = t_{n+1} - t_n$ are independent and identically distributed random variables. Let $t_{\rm tick}$ be the mean and $\Delta t_{\rm tick}$ be the standard deviation of the waiting time distribution. The resolution is defined by
\begin{equation}\label{resolution}
\nu_{\rm tick} = t_{\rm tick}^{-1},
\end{equation}
i.e. the average frequency of ticks. The accuracy is defined as the number of ticks, $N$, after which the total time uncertainty exceeds the mean time between ticks. This is a simple measure of the number of reliable ticks that the clock can produce. Since the uncertainty of $n$ independent and identically distributed ticks is equal to $\sqrt{n} \Delta t_{\rm tick}$, it follows that
\begin{equation}\label{accuracy}
N = \left (\frac{ t_{\rm tick}}{\Delta t_{\rm tick}}\right )^2.
\end{equation} 

A key question now is how the clock's performance depends on its energy consumption. The heat absorbed from the hot bath per tick is $Q_h = (d-1)\omega_h$. However, here it is important to note that the usefulness of a clock's output is not directly related to the energetic content of the ticks. In principle, a large proportion of the energy carried by each photon could be reused for another purpose, or simply dumped back into the hot bath, without affecting the clock's operation. We therefore quantify the clock's energy consumption by the heat dissipated into the cold bath per tick,
\begin{equation}\label{dissipated_power_clock}
Q_c = (d-1)\omega_c.
\end{equation}
This represents the minimum energy that is irreversibly expended in order to generate each tick. 

\begin{figure}
	\begin{minipage}{0.33\linewidth}
		\flushleft	\small (a)
				\includegraphics[width=\linewidth]{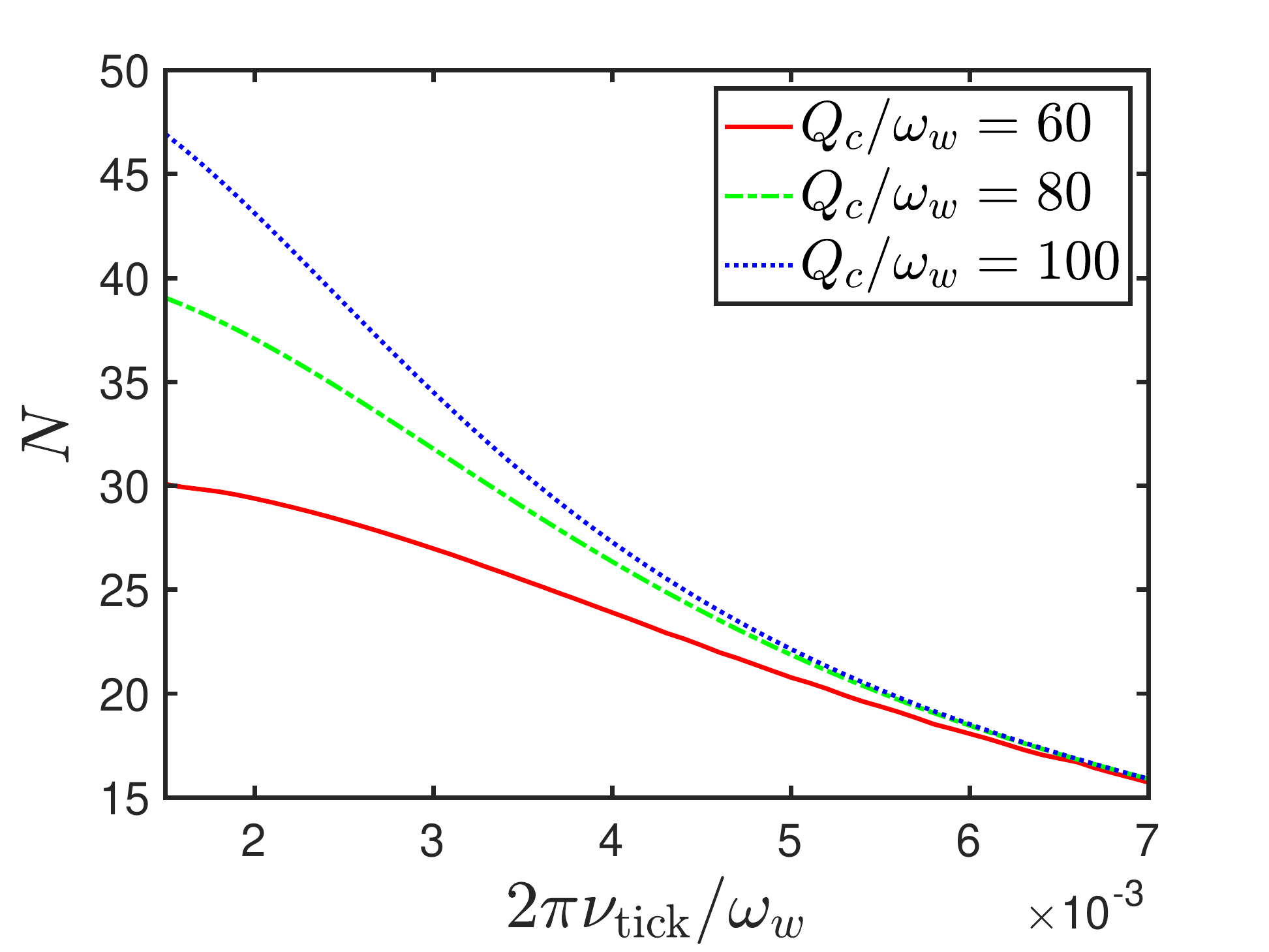}
	\end{minipage}
	\begin{minipage}{0.33\linewidth}
		\flushleft 	\small (b)
				\includegraphics[width=\linewidth]{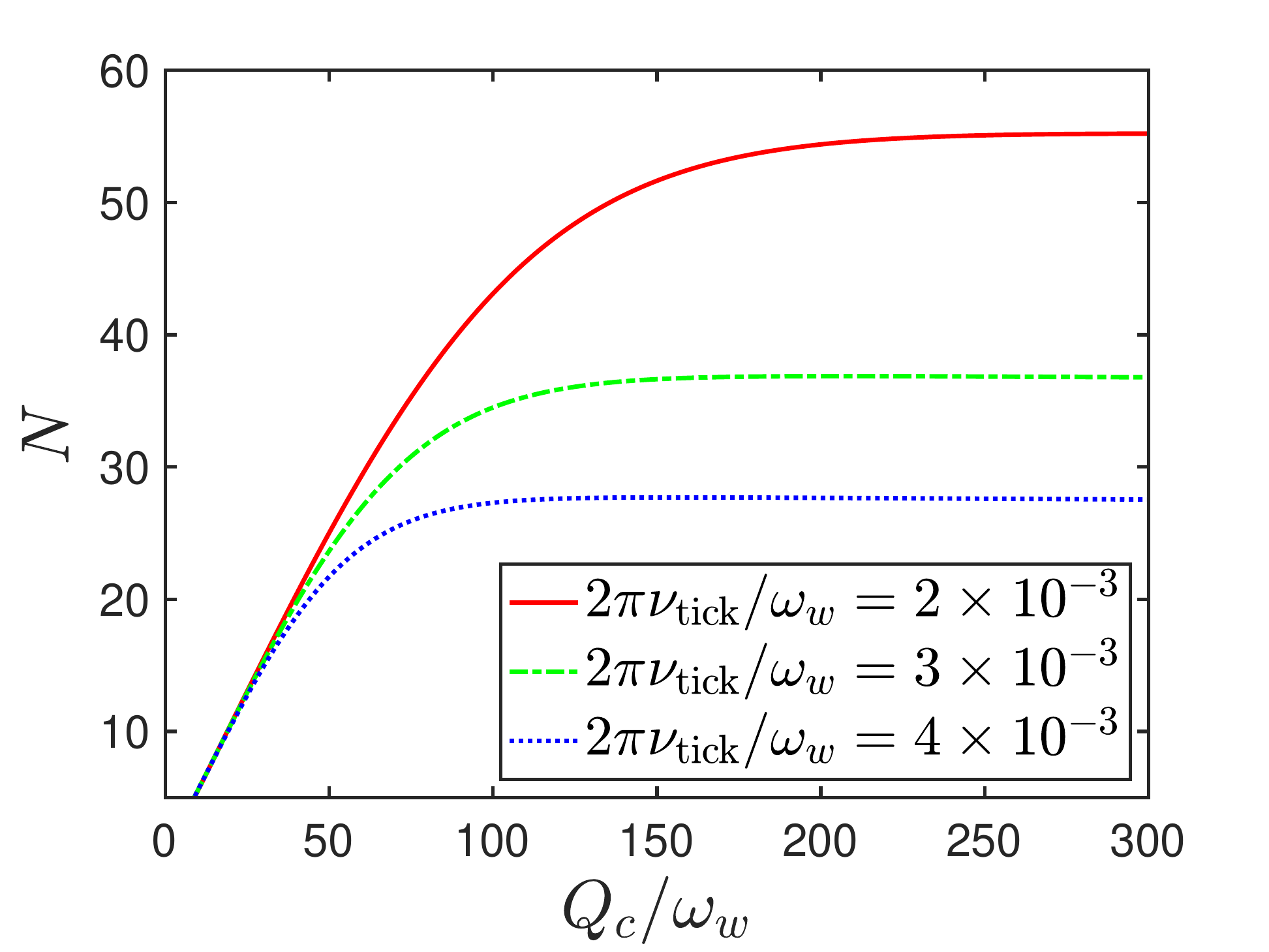}
	\end{minipage}
	\begin{minipage}{0.33\linewidth}
		\flushleft 	\small (c)
		\includegraphics[width=\linewidth]{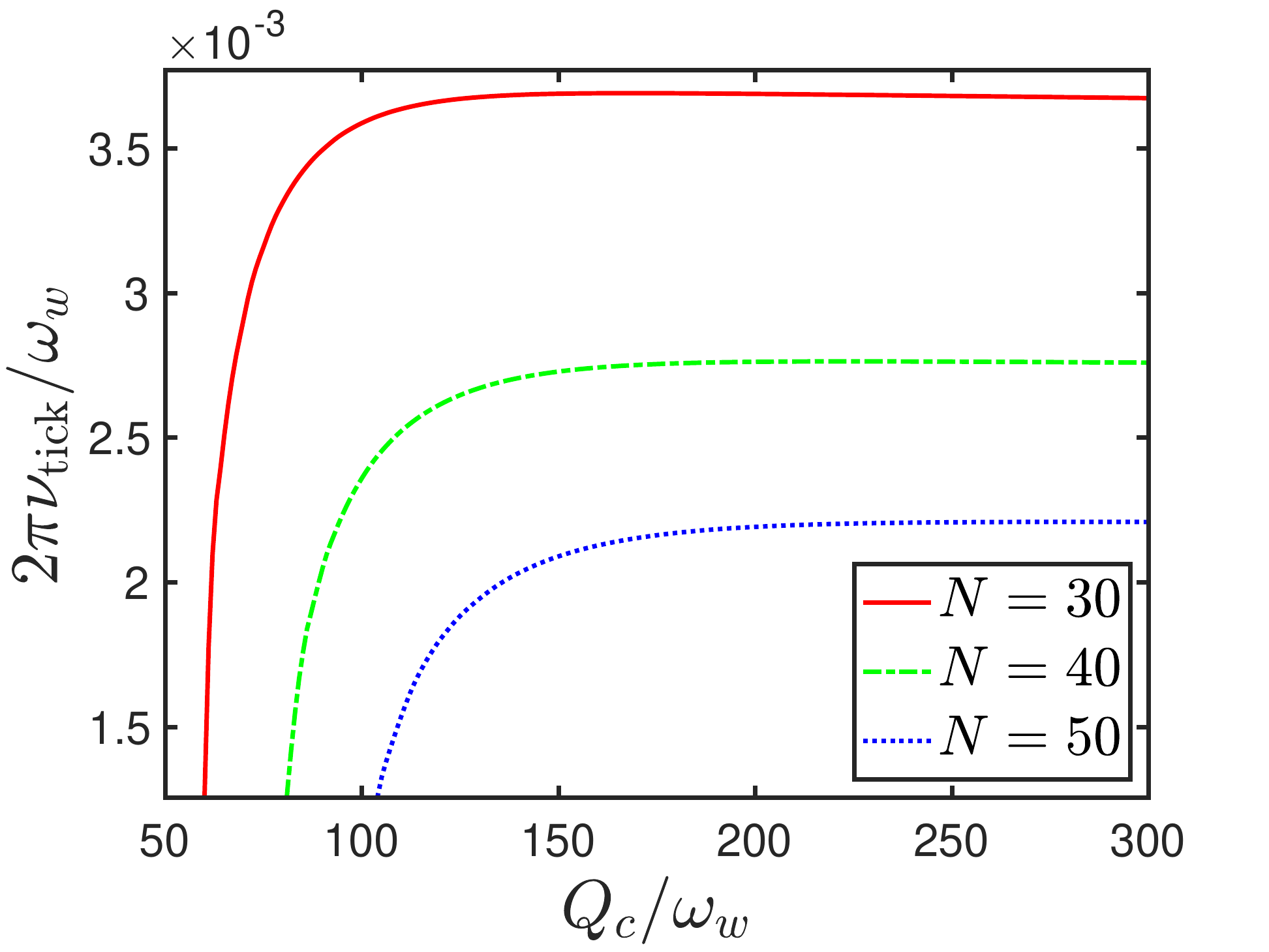}
	\end{minipage}
	\caption{Tradeoff between accuracy, resolution and dissipated power in a thermal absorption clock. The plots show: (a)~accuracy against resolution at fixed power consumption, (b) accuracy as a function of power consumption at fixed resolution and (c) resolution against power at fixed accuracy. Figure adapted from Ref.~\cite{Erker2017}. \label{fig:clock_tradeoff}}
\end{figure}

The relationship between power consumption and performance is illustrated in Fig.~\ref{fig:clock_tradeoff} for some examples. At any fixed power consumption, there is a tradeoff between accuracy and resolution: one may be improved only at the expense of the other, as shown in Fig.~\ref{fig:clock_tradeoff}(a). In order to increase the clock's accuracy without sacrificing resolution, or vice versa, it is necessary to consume more power. However, increasing the power consumption beyond a certain point offers no advantage due to the saturation behaviour demonstrated in Fig.~\ref{fig:clock_tradeoff}(b,c). 

An intuitive understanding of the relationship between accuracy and dissipated power can be gained by considering the limit of very low resolution, where the qubits thermalise quickly relative to the rate of energy transfer to the load. As discussed in Sec.~\ref{sec:three_body_eng}, this leads to an effective description of the load dynamics (in between each photon emission) as the biased random walk described by Eq.~\eqref{biased_random_walk}. It is thus straightforward to deduce the resolution from the mean time needed for the walker to take $d$ upward steps:
\begin{equation}\label{resolution_random_walk}
t_{\rm tick} = \frac{d}{\gamma_\uparrow- \gamma_\downarrow}.
\end{equation}
For large $d$, the relative uncertainty in each tick can be estimated from the relative uncertainty in the walker's position at time $t_{\rm tick}$, i.e. $\Delta t_{\rm tick}/t_{\rm tick} = \sigma(t_{\rm tick})/d$~\cite{Erker2017}. This yields the accuracy
\begin{equation}\label{accuracy_random_walk}
N = d \tanh\left (\frac{\Delta S_{\rm tick}}{2d}\right ),
\end{equation}
where $\Delta S_{\rm tick} = \beta_c Q_c - \beta_h Q_h$ is the entropy production per tick (assuming that the photon energy is not recycled). In particular, in the limit of very large dimension, $d\to \infty$, the accuracy is limited only by entropy production
\begin{equation}\label{accuracy_entropy_d_inf}
N\to \frac{\Delta S_{\rm tick}}{2}.
\end{equation}

Similar entropic constraints on timekeeping emerge from classical stochastic thermodynamics~\cite{Roldan2015,Barato2016}. Such constraints are closely connected to the recently-discovered thermodynamic uncertainty relations, which relate the fluctuations of currents (such as a stream of physical ticks) to the entropy produced in a classical Markovian process~\cite{Barato2015,Gingrich2016,Pietzonka2016,Pietzonka2017,Horowitz2017}. The extent to which such bounds also apply to quantum systems is only just beginning to be explored~\cite{Macieszczak2018,Guarnieri2019}. It is known, however, that quantum systems can violate the classical bounds in some cases, exhibiting reduced current fluctuations for a given entropy production~\cite{Brandner2018,Ptaszynski2018,Agarwalla2018}. This suggests that quantum clocks may be subject to less stringent thermodynamic limitations than classical ones. However, no clear quantum advantage appears in the present model because the dynamics of the load itself are essentially classical. That is, the load's reduced state lacks coherence in the energy eigenbasis and its evolution is well approximated by a classical stochastic process.

Many intriguing questions regarding the quantum thermodynamics of autonomous clocks remain largely unexplored. For example, alternative models could be devised in which quantum coherence plays a prominent role. This is important in order to connect these abstract results to practical, technological applications such as atomic clocks. Furthermore, it is necessary to quantify the thermodynamic cost of the measurement process itself~\cite{Jacobs2009,Jacobs2012,Abdelkhalek2016,Guryanova2018}, which represents an additional energy expenditure that was not considered in Ref.~\cite{Erker2017}. It would also be interesting to understand whether reservoirs featuring multiple conserved quantities~\cite{Guryanova2016,Halpern2016}, or even completely different kinds of passive resource states, might offer some advantage. More generally, the question of whether quantum and classical clocks have fundamentally different performance characteristics is an active topic of research~\cite{Woods2018a}.

\section{Summary and outlook}
\label{sec:summary}

In summary, studying autonomous thermal machines allows us to identify the ultimate thermodynamic cost of tasks such as cooling, work production and timekeeping in terms of heat, which is a universal and ubiquitous energy resource. Here we have focused on thermal absorption machines, i.e. continuous devices for which energy is the only conserved quantity. The functioning of these machines is intimately connected with the energy quantisation exhibited by small quantum systems. In particular, specific pairs of states with well-defined transition frequencies act as filters for the broadband thermal baths. This allows for the engineering of virtual temperatures that determine the operating regime of the machine, be it refrigerator, heat pump or engine. 

Roughly speaking, the optimal efficiency is obtained when only a small number of transition frequencies play a role in the machine's operation, as is typical in the weak-coupling regime. Given this constraint, the Hilbert-space dimensionality turns out to be a useful resource for reducing temperature in refrigerators and for achieving the best accuracies in clocks. On the other hand, strong interactions may boost power but typically at the expense of additional entropy production, which reduces efficiency. We also identified intrinsic heating in the load due to thermal fluctuations as a source of internal irreversibility that affects the performance of heat engines operating at finite energy scales, even in the weak-coupling regime. More broadly, quantum resources play a useful role in specific scenarios, such as single-shot cooling, but the question of whether quantum effects provide a generic advantage remains moot.

Much of the work on absorption machines thus far has focused on average quantities such as the mean heat currents. A natural next step is to consider the fluctuations of these currents and the corresponding fluctuations of efficiency. Indeed, in the context of autonomous clocks, such fluctuations are the primary performance indicator, being related to the accuracy as discussed in Sec.~\ref{sec:clocks}. The connection between efficiency, power and fluctuations has been investigated extensively for classical engines~\cite{Brandner2015,Shiraishi2016,Pietzonka2018,Holubec2018a} but less so in the quantum case~\cite{Guarnieri2019}. In the context of absorption machines, heat and power fluctuations have been respectively examined in the three-level refrigerator~\cite{Segal2018} and engine~\cite{Li2017}. However, we expect multipartite machines to exhibit even more interesting behaviour in this regard due to the essential role of coherence in energy transport. In particular, the potential of quantum systems to violate the classical thermodynamic uncertainty relations points towards the possibility of demonstrating unambiguous quantum advantages for absorption machines. 

This review focused explicitly on situations where the thermal baths couple weakly to the machine. However, one could also envisage absorption machines operating in regimes of strong system-reservoir coupling. In fact, strong correlations with the environment may enhance, or at least qualitatively modify, the performance of quantum heat engines~\cite{Gelbwaser-Klimovsky2015a,Strasberg2016,Xu2016,Newman2017,Perarnau-Llobet2018}. An accurate description of such problems requires non-perturbative methods~\cite{Jang2008,Nazir2009,Prior2010,Iles-Smith2014,Tamascelli2018,Strathearn2018}, which typically involve incorporating some or all of the environment explicitly into the description of the quantum system. Such procedures are, of course, fully consistent with our philosophy of autonomy, since they permit careful accounting of the effects of energy exchange between the quantum system and its surroundings.

Finally, we mention an entirely different class of absorption machines designed to generate quantum resources, such as coherence and entanglement, using only heat flows. For example, simple two-body machines operating between two heat reservoirs at different temperatures suffice to generate maximally entangled states~\cite{Brask2015a,Tavakoli2018}. The creation~\cite{Guarnieri2018a} and amplification~\cite{Manzano2019} of coherence using thermal baths has also been discussed. Such machines are indisputably quantum by nature since their successful operation is classically impossible by definition. One of the most intriguing aspects of the interplay between quantum information and thermodynamics is the appearance of additional resources besides work~\cite{Chitambar2019}, such as squeezed or non-Gaussian states and many other examples. Designing absorption machines to generate such resources opens up completely new possibilities for heat conversion at the smallest scales. 

\section*{Acknowledgements}
Useful discussions with John Goold during the preparation of the manuscript are gratefully acknowledged. This work was funded by the European Research Council under the European Union's Horizon 2020 research and innovation program (grant agreement 758403). 

\section*{About the author}
Mark Mitchison is a theoretical physicist based at Trinity College Dublin. Mark wrote his PhD thesis on atomic-scale absorption refrigerators and thermometers at Imperial College London in 2016. He then worked for two years at Universit\"at Ulm before moving to Dublin in 2018. His research focuses on open quantum systems and their applications as thermal machines and as platforms for investigating non-equilibrium physics.

\let\c\cedilla
\bibliographystyle{apsrev4-1}
\bibliography{article}

\end{document}